\documentclass[final,1p,times,twocolumn]{elsarticle}

\usepackage{graphics}

\usepackage{amssymb}

\journal{Comptes Rendus Physique}

\begin{document}

\begin{frontmatter}



\title{Antiferromagnetism and Superconductivity in Cerium based Heavy Fermion Compounds}


\author{Georg Knebel}
\author{Dai Aoki}
\author{Jacques Flouquet}

\address{SPSMS, UMR-E CEA / UJF-Grenoble 1, INAC, 38054 Grenoble, France}

\begin{abstract}
The study of competing ground states is a central issue in condensed matter physic. In this article we will discuss the interplay of antiferromagnetic order and unconventional superconductivity in Ce based heavy-fermion compounds. In all discussed examples superconductivity appears at the border of magnetic order.  Special focus is given on the pressure--temperature--magnetic field phase diagram of CeRhIn$_5$ and CeCoIn$_5$ which allows to discuss  microscopic coexistence of magnetic order and superconductivity in detail. A striking point is the similarity of the phase diagram of different classes of strongly correlated systems which is discussed briefly. The recently discovered non-centrosymmetric superconductors will open a new access with the possible mixing of odd and even parity pairing. 

\end{abstract}
\end{frontmatter}


\section{Introduction}
\label{introduction}

The interplay of rivaling ground states and the appearance of unconventional superconductivity close to quantum critical points is a central issue in the physics of strongly correlated electron systems. This includes the search for new superconductors but also the understanding of the superconducting pairing mechanism. Ideal model systems for these studies are metallic heavy-fermion systems \cite{Flouquet2005}, but includes also high-$T_c$ cuprates \cite{Sachdev2010}, organic charge transfer salts \cite{Bourbonnais2008}, and the recently discovered iron pnictide and chalcogenide superconductors \cite{Paglione2010}. In all these systems the superconducting phases exist at the border of rivaling ground states such as a magnetically ordered state and a paramagnetic state. In heavy-fermion systems both coexistence of antiferromagnetism (AF) or ferromagnetism (FM) with superconductivity has been observed and have been intensively studied. 

\begin{figure}[t]
\begin{center}
\includegraphics[width=0.6\hsize,clip]{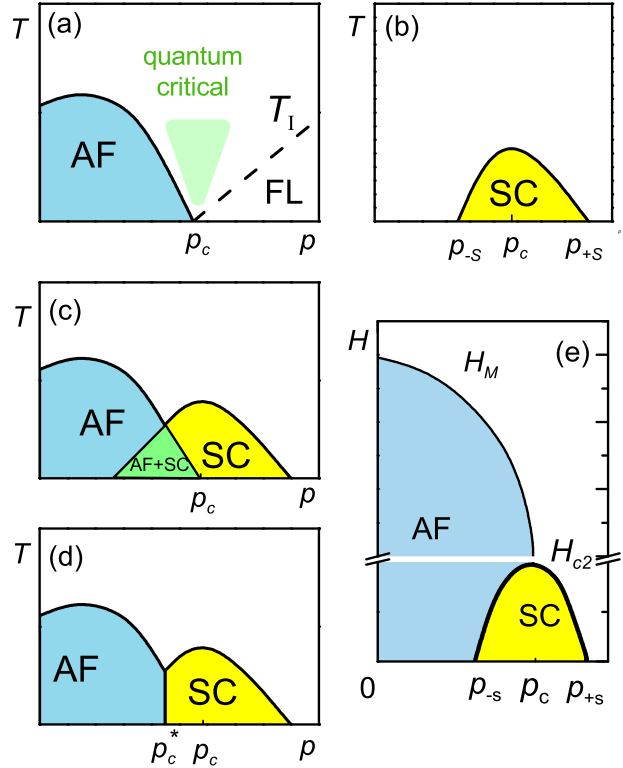}%
\caption{\label{f1} Schematic phase diagram of heavy fermion compounds indicating the interplay of antiferromagnetism (AF) and superconductivity (SC). (a) Temperature pressure phase diagram in absence of superconductivity. Close to the critical pressure $p_c$ quantum criticality gets important and the magnetic order and the Fermi liquid regime collapses. (b) Superconducting phase without magnetic order. (c) Coexistence of AF and SC, (d) first order transition and separation of magnetism and superconductivity. (e) Magnetic field -- pressure phase diagram at $T = 0$ with the collapse of $H_M$ at the critical pressure and its interplay with the superconducting dome around $p_c$. }
\end{center} 
\end{figure}

Intermetallic lanthanide or actinide compounds allows detailed studies of the interplay of different orders. The strong Coulomb repulsion in the 4$f$ of 5$f$ shells and the strong hybridization with the conduction electrons give rise to the formation of heavy quasiparticles with a strong enhancement of the electron mass $m^\star$ up to 100 times higher than that of a usual metal. The strong electronic interactions can lead to a variety of different ground states: paramagnetic or magnetically ordered, coupled or not with superconductivity. Due to the very large electronic Gr\"uneisen parameter the ground state of the heavy electron systems is very sensitive to external control parameters like pressure $(p)$, doping, or magnetic field $(H)$ and it can be tuned continuously from one ordered state to the other. By suppressing the ordering temperature to $T=0$ a quantum phase transition (QPT) can be induced by pressure. If a second order phase transition vanishes continuously to $T=0$ as function of a non-thermal control parameter a so-called quantum critical point (QCP) can be achieved. In principle the heavy fermion state is that of a renormalized Fermi liquid with  the linear $\gamma T$ term of the specific heat $C$  and the Pauli susceptibility $\chi_0$ strongly enhanced by a factor of 100 to 1000 in comparison to a classical metal. Furthermore the $A$ coefficient of the electrical resistivity $\rho (T) = \rho_0 + A T^2$ is very large and related to the specific heat coefficient $\gamma = C/T$ by the so-called Kadowaki-Woods relation $A/\gamma ^2 = {const}$ due to the strong local character of the magnetic fluctuations \cite{Kadowaki1986}. However, at a finite region around the critical value of the control parameter significant deviations from the Fermi-liquid behavior are observed experimentally with a strongly increasing specific heat coefficient $C/T$ and a non quadratic temperature dependence of the resistivity to lowest temperatures. The origin of the non-Fermi liquid behavior is directly linked to the diverging coherence length of critical fluctuations at the critical point at $T=0$. A recent review on quantum phase transitions in heavy fermion systems is given in Refs.~\cite{Loehneysen2007,Gegenwart2008}. 

In difference to a classical phase transition at finite temperature where critical fluctuations are restricted on a small fraction of a reduced temperature scale, such an energy scale does not exist at a quantum phase transition and the system is governed by quantum fluctuations. However, this standard picture of a spin fluctuation driven transition from a magnetically ordered system to a paramagnetic system is not universal. E.g.~the collapse of the magnetic order in the archetype non-Fermi liquid systems CeCu$_{6-x}$Au$_x$ or YbRh$_2$Si$_2$ cannot be described in the frame of spin-fluctuation theory. One recent theoretical concept is the so-called Kondo-breakdown scenario \cite{Si2010}. Here, the corresponding transition from the paramagnetic to a magnetic ground state is referred to as local criticality where the antiferromagnetic transition is accompanied by a partial localization of the f-electrons and a significant change of the Fermi surface volume is expected to appear at the critical pressure \cite{Si2010}. Other theoretical scenarios include an selective Mott transition of the $f$ electrons \cite{Pepin2007}, and for YbRh$_2$Si$_2$ it has been shown that many properties of the field driven quantum critical point in that system can be described by assuming a Zeeman-driven Lifshitz transition of narrow quasi-particle bands \cite{Hackl2011}. Superconductivity has not been observed in these systems. 

While previous approaches discuss the spin degrees of freedom,  critical valence fluctuations give rise to quantum criticality too, see Ref.~\citealp{Miyake2007}). There it is shown that not only critical magnetic fluctuations may drive the system to superconductivity, but also critical valence fluctuations. This model has been basically developed after the observation of a second superconducting regime in CeCu$_2$Si$_2$ and CeCu$_2$Ge$_2$ under high pressure \cite{Jaccard1999, Holmes2004, Yuan2003}. Very often both the magnetic and valence criticality are coupled, and both channels will contribute to the Cooper pairing.

In Fig.\ref{f1}(a) the pressure--temperature phase diagram of a heavy-fermion systems in a conventional spin fluctuation picture is plotted. Antiferromagnetic order vanishes at a quantum critical point and Fermi-liquid behavior is recovered on the paramagnetic side \cite{Millis1993, Moriya1995}. 
Figure \ref{f1}(b) shows the phase diagram of a superconductor, and the superconducting state is expected to exist over a dome in the pressure-temperature plane. In heavy-fermion systems these two phase diagram appears to be combined and a dome like unconventional superconducting state appears in the vicinity to the QPT where the quantum fluctuations diverge. These critical fluctuations are responsible for the attractive pairing interaction for the unconventional superconductivity \cite{Monthoux2007}. 
In Fig.~\ref{f1}(c-d) we sketch the phase diagrams with (i) a coexistence regime of antiferromagnetic order and superconductivity AF+SC, with four second order transition lines coming together at a tetra-critical point at $T_N = T_c$ for a critical pressure $p_c^\star$, or (ii) a first order transition between the antiferromagnetic phase and superconductivity which will also appear at a critical pressure $p_c^\star$. In that case, phase separation between magnetic non-superconducting and paramagnetic superconducting phases may appear without any homogeneous coexistence. The application of a magnetic field at $T=0$ will lead to the phase diagram sketched in Fig.~\ref{f1}(d).  The important novelty will be the interplay of antiferromagnetism and superconductivity in the vortex state below $p_c$. Again, the coexistence of magnetism and superconductivity will appear below the critical pressure.

In the following we want to concentrate on Ce-based heavy-fermion compounds. The ground-state properties of these compounds are, in a simple picture, determined by the competition of the onsite Kondo interaction  (where due to the exchange coupling  of the local 4$f$ spin to the conduction electrons the magnetic moment is screened and thus a paramagnetic ground state is formed)  and the inter-site RKKY interactions (which gives rise to a magnetically ordered state). The competition of both interactions has been discussed in the so-called Doniach model \cite{Doniach1977}. Both interactions depend critically on the hybridization strength of the 4$f$ states and the conduction states which can be easily modified by applying hydrostatic pressure. Thus, it is possible to tune an antiferromagnetic heavy-fermion systems from a magnetically ordered state to a paramagnetic state by applying pressure as shown schematically in Fig.~\ref{f1}.  

Heavy-fermion superconductivity has been first observed in CeCu$_2$Si$_2$ \cite{Steglich1979}. All superconductors discovered before had been conventional (phonon-mediated) superconductors and small amounts of magnetic impurities suppresses superconductivity in a conventional superconductor (see e.g. Ref.~\citealp{Maple1973}). On the contrary, in CeCu$_2$Si$_2$ the lattice is formed of Ce$^{3+}$ Kondo ions at regular lattice sites and small amounts of doping with non-magnetic impurities suppresses the superconducting state \cite{Spille1983}. At that time CeCu$_2$Si$_2$ was thought to be a paramagnetic heavy-fermion superconductor, but ten years later observations of tiny antiferromagnetic order have been reported \cite{Nakamura1988, Uemura1989}. In the 80th and beginning of the 90th superconductivity has been found in different U-based heavy-fermion compounds:  UBe$_{13}$ \cite{Ott1983}, UPt$_3$ \cite{Stewart1984},URu$_2$Si$_2$ \cite{Palstra1985}, UNi$_2$Al$_3$ \cite{Geibel1991b}, and UPd$_2$Al$_3$ \cite{Geibel1991a}. The development of high pressure experiments and the progress in sample quality led to the observation of pressure induced superconductivity in several Ce-based heavy fermion compounds close to their magnetic instability. The first example has been CeCu$_2$Ge$_2$ which is isoelectronic to CeCu$_2$Si$_2$ \cite{Jaccard1992}, and soon after  CeRh$_2$Si$_2$ \cite{Movshovich1996}, CeIn$_3$ and CePd$_2$Si$_2$ \cite{Mathur1998,Grosche1996}. These compounds are antiferromagnetically ordered at ambient pressure and superconductivity appears
only under high pressure and at very low temperature ($T_c < 600$ mK), which makes a detailed analysis of the
competition of both phenomena experimentally very difficult. A breakthrough was the discovery of superconductivity in the Ce$T$In$_5$ ($T=$Co, Rh, Ir) compounds \cite{Hegger2000, Petrovic2001, Petrovic2001b}. In this so-called Ce-115 family of compounds it is possible to perform precise measurements of the magnetic as well as of the superconducting properties and to study their interaction thanks to their high superconducting transition temperature of $T_c \approx 2$ K .

The superconductivity in the recently discovered systems with  a non-centrosymmetric crystal structure like CePt$_3$Si \cite{Bauer2004} and the Ce-113 family Ce$TX_3$ ($T$= Co, Rh, Ir; $X$= Si, Ge) \cite{Settai2007} show novel superconducting pairing mechanisms and very exciting properties, as e.g.~the very huge upper critical fields close to the critical pressure \cite{Settai2008}. 

Up to now more that 30 different heavy-fermion compounds have been observed which show superconductivity at ambient or high pressure \cite{Pfleiderer2009}. In this article we want to discuss selected examples: (i) CeCu$_2$Si$_2$ and CeCu$_2$Ge$_2$ showing two different superconducting domes under pressure with the one at low pressure  being connected to a magnetic instability and the high pressure dome to critical valence fluctuations \cite{Holmes2004}. (ii) We will review the examples of CePd$_2$Si$_2$ and CeIn$_3$ which show rather low superconducting transition temperatures. (iii) The Ce-115 family with the special focus on CeRhIn$_5$ and CeCoIn$_5$ will be reviewed, and (iv) finally we will shortly discuss the case of the non-centrosymmetric compound CeIrSi$_3$ studied in Grenoble thanks our collaboration with Osaka University.

\section{CeCu$_2$Si$_2$}

\begin{figure}[t]
\begin{center}
\includegraphics[width=0.6\hsize,clip]{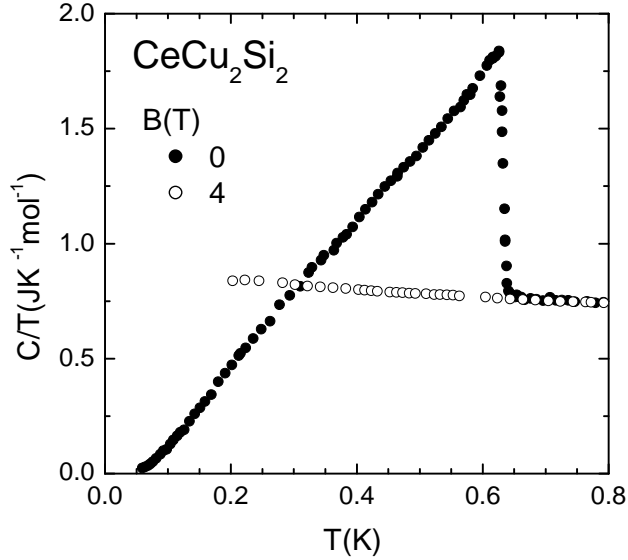}%
\caption{\label{CeCu2Si2} Specific heat divided by temperature of CeCu$_2$Si$_2$ in zero field and at 4~T (from \cite{Steglich1996}) }
\end{center} 
\end{figure}

\begin{figure}[t]
\begin{center}
\includegraphics[width=0.6\hsize,clip]{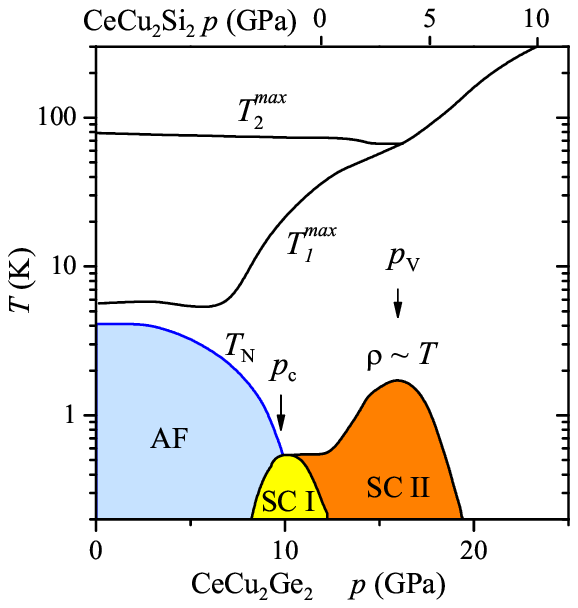}%
\caption{\label{CeCu2Si2_pressure} High pressure phase diagram of CeCu$_2$Ge2$_2$ and CeCu$_2$Si$_2$ showing  two superconducting domes, SC I and SC II. SC I appears at the magnetic quantum critical point $p_c$, SC II is centered at the critical valence transition $p_V$. $T_1^{max}$ and $T_2^{max}$ indicate the maxima of the resistivity corresponding to the Kondo temperature and to the Kondo temperature of the full multiplet. (adapted from Ref.~\cite{Holmes2004}. Experiments on Ge doped crystals of CeCu$_2$Si$_2$ show the separation of both domes \cite{Yuan2003}. }
\end{center} 
\end{figure}

The discovery of the first heavy-fermion superconductor, CeCu$_2$Si$_2$ in 1979 \cite{Steglich1979}, which also has been the first case of superconductivity in strongly correlated electron systems, was quite a surprise (see Fig.~\ref{CeCu2Si2}) as just above $T_c\approx 600$~mK the magnetic entropy coming from the spin of the Ce atoms is still large ($\sim 0.1$~R$\log 2$) and thus, at first glance, local spin fluctuations acting as individual depairing Kondo centers are considered to be a strong pair breaking mechanism. It took almost three decades to clarify the case of CeCu$_2$Si$_2$ as its superconducting properties are very sensitive to the sample stoichiometry \cite{Modler1995}. Now it is well established that CeCu$_2$Si$_2$ at ambient pressure is close to an antiferromagnetic instability. Tiny differences in the composition can induce antiferromagnetism. However, the same incommensurate wave vector \boldmath $Q_{af}$ \unboldmath $=(0,215\; 0,215\; 0,53)$ was observed associated to strong magnetic fluctuations or to long range magnetic order with the magnetic ordering vector being connected to Fermi surface nesting \cite{Stockert2004}. Extensive NMR experiments have already demonstrated the strong interplay between antiferromagnetism and superconductivity \cite{Ishida1999, Kitaoka2001} in CeCu$_2$Si$_2$. However, by contrast to the Ce-115 series (see below), superconductivity and magnetic order do not coexist, but repeal each other, i.e.~superconductivity seems to push out the antiferromagnetism \cite{Faulhaber2007} which may be forced to disappear via a first order transition.

High pressure experiments performed on CeCu$_2$Si$_2$, Ge doped, and pure CeCu$_2$Ge$_2$ have shown that the $(T,p)$ phase diagrams of these systems have two superconducting domes \cite{Jaccard1992, Jaccard1999, Yuan2003, Holmes2004} as shown in Fig.~\ref{CeCu2Si2_pressure}. SC I is governed by magnetic fluctuations and SC II is caused by valence fluctuations (see Fig.~\ref{CeCu2Si2_pressure}). Thanks to studies on CeCu$_2$Ge$_2$ which is at ambient pressure an antiferromagnet, there is no doubt that the appearance of superconductivity (SC I) is linked to the collapse of magnetism. 

The disappearance of the  crystal field splitting, the large linear $T$ term of the resistivity, an anomaly in the pressure dependence of the thermoelectric power, and the concomitant strong decrease of the Kadowaki Woods ratio $A/\gamma^2$ \cite{Jaccard1999, Holmes2004} indicate that for $p \sim p_V$, where the second superconducting dome SC II is centered, the Ce ions enter in a new weakly  correlated electronic phase. It is characterized by a sharp crossover from a quasi-trivalent state of the Ce ions to an intermediate valence state which is associated with the loss of the crystal field effect, e.g.~to a change of the degeneracy from 2 to 6 being the degeneracy of the full multiplet. Even if the occupation number of the $4f$ shell drops only from $n_f \sim 0.98$ to $n_f \sim 0.9$ at $p \sim p_V$ the consequences on the spin dynamic will be large as its local temperature ($T_K \sim 1/(1-n_f)$) depends critically on small variations of $n_f$. Experimentally it has never been demonstrated by  high energy spectroscopy that an abrupt valence change occurs at $p_V$ (see e.g.~recent work using resonant X-ray scattering \cite{Rueff2011}). It may correspond only to a smooth crossover \cite{Watanabe2010}. However there is microscopic evidence by NQR experiments that the electrical field gradient and the crystal field splitting change around $p=4$~GPa \cite{Fujiwara2008}. At least it is well established that in CeCu$_2$Si$_2$  spin and valence criticality are well separated \cite{Yuan2003}. In the following examples of CeIn$_3$ and CePd$_2$Si$_2$, they cannot be clearly distinguished. 

Up to now, only a few orbits with light carriers have been detected by quantum oscillations in CeCu$_2$Si$_2$ due to the combined effects of the large masses and and single crystals with rather poor residual resistivity ratios \cite{Springford1991}. As will be shown below, the high quality of the crystals of the Ce-115 family gives the opportunity to correlate the occurrence of superconductivity with the Fermi surface topology. 

Finally, concerning the superconducting order parameter, singlet pairing  with nodes of the gap occurs, but there is no microscopic evidence of the exact location of line or point nodes. Recent inelastic neutron experiments indicates also the occurrence of superconducting resonance characteristic of singlet pairing at the antiferromagnetic wave vector \boldmath $Q_{af}$ \unboldmath \cite{Stockert2011} but the signal is rather broad and extends ten times the gap value whereas in  CeCoIn$_5$ a much sharper, resolution-limited resonance has been found \cite{Stock2008} (see below chapter \ref{UCS_CeCoIn5}).

\section{Low $T_c$ compounds :  CePd$_2$Si$_2$ and CeIn$_3$}

\begin{figure}[t]
\begin{center}
\includegraphics[width=1\hsize,clip]{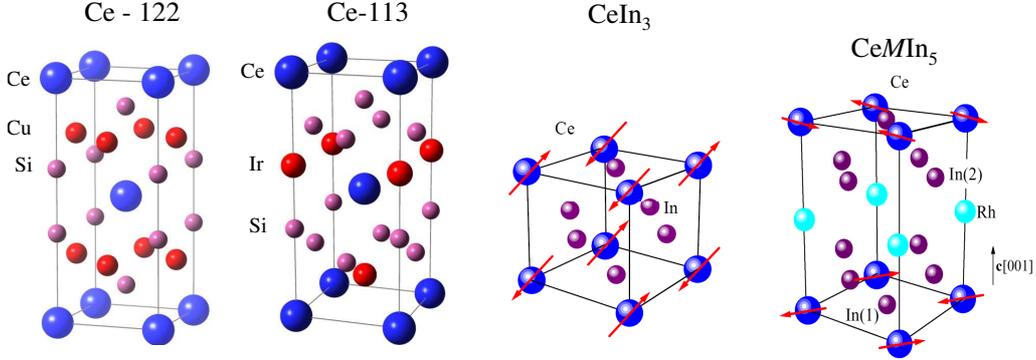}%
\caption{\label{crystal_structure} From lest to right : ThCr$_2$Si$_2$ crystal structure of Ce-122, the BaNiSn$_3$ structure of non centrosymmetric Ce-113 systems, the cubic structure of CeIn$_3$, and HoCoGa$_5$ structure of the Ce-115, and (right).  }
\end{center} 
\end{figure}

\begin{figure}[t]
\begin{center}
\includegraphics[width=0.6\hsize,clip]{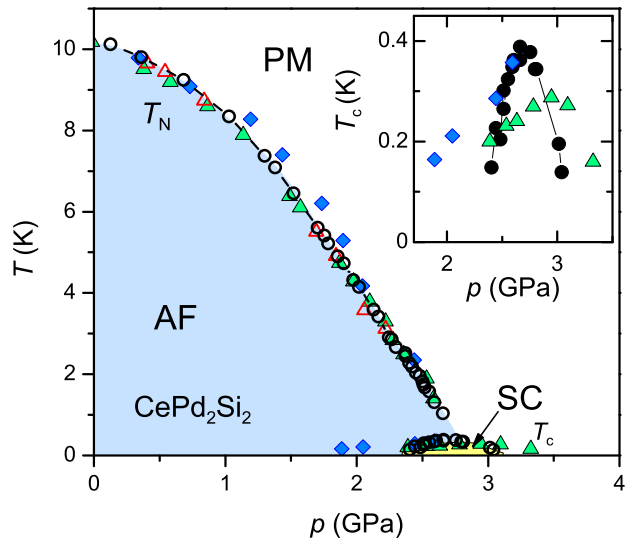}%
\caption{\label{PD_CePd2Si2} Pressure--temperature phase diagram of CePd$_2$Si$_2$ under hydrostatic conditions. The inset shows a zoom on the superconducting transitions. Circles are taken from Ref.~\citealp{Mathur1998}, diamonds from  Ref.~\citealp{Sheikin2001}, and triangles from Ref.~\citealp{Demuer2001}. In Ref.~\citealp{Mathur1998} the superconducting transition temperature $T_c$ is defined by the midpoint of the superconducting transition, while in refs.~\citealp{Sheikin2001,Demuer2001}, $T_c$ has been defined by the onset of the transition and a complete superconducting transition has been only observed close to the maximum of $T_c$.  }
\end{center} 
\end{figure}

\subsection{CePd$_2$Si$_2$}

CePd$_2$Si$_2$ crystallizes in the same tetragonal structure than CeCu$_2$Si$_2$ (see Fig.~\ref{crystal_structure}). At ambient pressure antiferromagnetic order develops below $T_N = 10.2$~K with an propagation vector \boldmath $Q$ \unboldmath $= (1/2, 1/2, 0)$ and with the magnetic moments oriented along [110] \cite{Grier1984}. The ordered magnetic moment $M_0 = 0.62 \mu_B$ is slightly reduced compared to the free Ce$^{3+}$ momentum in the crystal electric field. By inelastic neutron spectroscopy, Kondo-type spin fluctuations have been observed in the paramagnetic state above $T_N$ which coexist with spin-wave excitations below $T_N$ \cite{vanDijk2000}. 
The Kondo temperature $T_K \approx 10$ K $\approx T_N$ has been estimated from neutron scattering and from NMR measurements \cite{vanDijk2000,Kawasaki1998}. 
The Sommerfeld coefficient of the specific heat $\gamma = 125$~mJ/moleK$^2$  is slightly enhanced \cite{Besnus1991} and the resistivity at low temperatures can be well fitted with a Fermi liquid term $AT^2$ and an additional exponential term taking into account the spin-wave scattering \cite{Raymond2000}. Quantum oscillation experiments and the comparison to band-structure calculations implies that the 4$f$ electrons are itinerant rather than localized inside the magnetically ordered state at ambient pressure \cite{Sheikin2003}, even if the magnetic moment is rather large. Thus, CePd$_2$Si$_2$ is a moderate heavy-fermion systems and had been a very promising candidate to study the pressure induce quantum critical point. 

Pressure induced superconductivity has been first reported by Grosche et al. \cite{Grosche1996}. The pressure-temperature phase diagram of CePd$_2$Si$_2$ as determined from resistivity experiments by different authors is shown in Fig.~\ref{PD_CePd2Si2}. The experiments are performed under hydrostatic pressure conditions, either a liquid pressure transmitting medium in a large volume high pressure cell \cite{Mathur1998, Sheikin2001}, either in a diamond anvil cell with Helium as pressure medium. Experiments in Bridgman cell had shown a much larger superconducting region \cite{Raymond2000, Raymond1999}. The effect of non-hydrostatic pressure conditions in this tetragonal systems have been nicely demonstrated by careful experiments with weak uniaxial stress in a Bridgman type anvil cell.  It has been shown that a stress applied along the $c$ axis shifts the collapse of the magnetic transition and the superconducting transition to much higher pressures \cite{Demuer2002}, implying the need of excellent pressure conditions. 

Under hydrostatic pressure the magnetic transition temperature $T_N$ is monotonously suppressed and $T_N$ extrapolates to zero near $p_c = 2.8$~GPa. Remarkably, the pressure dependence $T_N (p)$ above $p=1.5$~GPa is linear which is not consistent with the predictions of the conventional spin-fluctuation model of a three dimensional antiferromagnet \cite{Millis1993, Moriya2003} and has been interpreted as indication of two dimensional fluctuations. As the magnetic structure of antiferromagnetically coupled ferromagnetic planes is stabilized by the  in-plane interactions only \cite{Villain1959}, such lower dimensionality could be realized. NMR experiments under high pressure may clarify this point. This interpretation has been supported by the temperature dependence of the resistivity, which shows close to the critical pressure $p_c = 2.8$~GPa strong deviations from Fermi-liquid behavior and a quasi-linear temperature dependence  $\rho (T) \propto T^n$ with $n \approx 1.2-1.3$ in a broad temperature range up to 40 K\cite{Mathur1998, Sheikin2001}. This non-Fermi liquid temperature dependence has been also observed under high magnetic field, at least up to $H = 6$ T \cite{Sheikin2001}. Another mark of the critical pressure is given by the strong enhancement of the $A$ coefficient of the resistivity at $p_c$.

Magnetic neutron diffraction experiments have been performed up to 2.45 GPa and a nearly linear relation between the magnetic ordering temperature and the staggered magnetization as a function of pressure has been found \cite{Kernavanois2005}, which supports an itinerant picture for the magnetism in CePd$_2$Si$_2$, in agreement with the quantum oscillation experiments at $p=0$. 

Superconductivity appears only on a small dome around the critical pressure $p_c \approx 2.8$ GPa (see inset of Fig. \ref{PD_CePd2Si2}). The phase diagram has been drawn only by resistivity, and away from the critical pressure the observed transition is not complete. Bulk nature of superconductivity at the critical pressure has been proved by ac calorimetry \cite{Demuer2002} where a specific heat anomaly has been detected at a temperature which coincides with the vanishing of resistivity. The upper critical field $H_{c2}$ and the initial slope $dH_{c2}/dT$ at $T_c$ are large and anisotropic ($H_{c2}^a (0) = 0.7$ T with $dH_{c2}^a/dT = -12.7$~T/K and $H_{c2}^c (0) = 1.3$~T with $dH_{c2}^c/dT = -16$~T/K) \cite{Sheikin2001} which indicates the pairing of heavy electrons. From these values the upper limit of the superconducting coherence length $\xi_0$ can be estimated and yielding values of 300~\AA\ and 230~\AA\ for the field along $a$ and $c$ axis, respectively. For both directions $\xi_0$ is much smaller than the estimated mean free path $\ell$ deduced from the residual resistivity $\rho_0 \approx 1 \mu\Omega$cm, indicating a clean limit. 

No experiments have been performed to determine the superconducting order parameter. The main problem for more detailed studies of the competition of antiferromagnetism and superconductivity in CePd$_2$Si$_2$ are the (i) sensitivity to non-hydrostatic pressure conditions, and (ii) the appearance of bulk superconductivity only at very low temperatures $T_c<400$~mK and on a small pressure dome centered at $p_c \approx 2.8$~GPa.

\subsection{CeIn$_3$}

\begin{figure}[t]
\begin{center}
\includegraphics[width=0.6\hsize,clip]{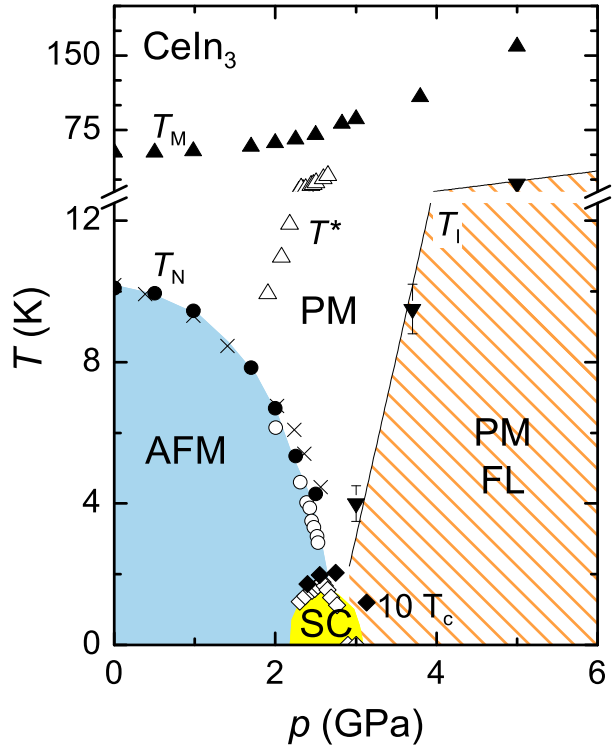}%
\caption{\label{PD_CeIn3} Pressure--temperature phase diagram of CeIn$_3$. At $p_c \approx 2.6$ GPa the antiferromagnetic order is suppressed and a paramagnetic Fermi liquid state is achieved for $p>p_c$ (taken from Ref.~\citealp{Knebel2001}). $T_M$ gives the high temperature maximum of the resistivity which is a measure for the crystal field splitting \cite{Knafo2003}. Open triangles indicate the temperature $T^\star$ where the spin lattice relaxation rate $1/T_1$ starts to decrease and gives a rough estimation of the Kondo temperature (taken from Ref.~\citealp{Kawasaki2008}). $T_1$ indicated the crossover to a Fermi liquid regime. 
} 
\end{center} 
\end{figure}

The Kondo-lattice compound CeIn$_3$ crystallizes in the simple cubic AuCu$_3$-type structure (see Fig.~\ref{crystal_structure}) and thus the pressure--temperature phase diagram is much less sensitive to different pressure conditions. Magnetic order appears in CeIn$_3$ below $T=10.1$ K and the magnetic structure is  simple type-II antiferromagnetic \cite{Benoit1980,Lawrence1980}. The Ce moments are aligned antiferromagnetically in adjacent (111) ferromagnetic planes. The ordered magnetic moment $M_0 = 0.5 \mu_B$ is reduced by comparison with the saturation moment of the $\Gamma_7$ doublet.  Like in CePd$_2$Si$_2$ inelastic neutron scattering on single crystals show quasi-elastic Kondo-type spin fluctuations and a broadened crystal field excitation above $T_N$ which coexists with well defined spin-wave excitations below $T_N$ in the antiferromagnetic state \cite{Knafo2003}. The crystal-field splitting between the low lying doublet and the $\Gamma_8$ quartet is about 125~K.

\begin{figure}[t]
\begin{center}
\includegraphics[width=0.5\hsize,clip]{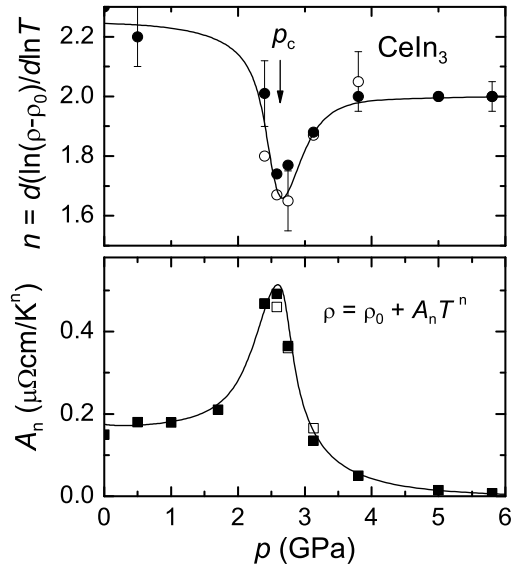}%
\caption{\label{CeIn3_AvsP} Pressure dependence of the exponent $n$ and the coefficient $A$ of the resistivity $\rho = \rho_0 + A_nT^n$(taken from Ref.~\citealp{Knebel2001}). 
} 
\end{center} 
\end{figure}

The pressure-temperature phase diagram of CeIn$_3$ shown in Fig.~\ref{PD_CeIn3} has been studied by resistivity \cite{Walker1997,Knebel2001}, ac calorimetry \cite{Knebel2002b}, NQR experiments coupled with ac susceptibility \cite{Kawasaki2002,Kawasaki2008}, neutron diffraction \cite{Flouquet1990}, de Haas van Alphen (dHvA) experiments \cite{Settai2005,Endo2005} and penetration depth studies \cite{Purcell2009}. The N\'{e}el temperature is monotonously suppressed under pressure and $T_N$ extrapolates to zero at $p_c \approx 2.6$~GPa. By contrast to CePd$_2$Si$_2$ the phase diagram is rather robust and no significant differences have been observed by changing pressure conditions. The exact location of the critical pressure $p_c$ varies from $p_c = 2.46$~GPa to $p_c = 2.65$~GPa. The temperature dependence of the resistivity shows in the antiferromagnetic ordered regime a stronger than $T^2$ dependence due to the presence of magnon scattering (see Fig.~\ref{CeIn3_AvsP}). Above the critical pressure a paramagnetic Fermi liquid ground state appears and the range of the existence of this regime increases linearly with pressure. At the critical pressure, a $T^{3/2}$ temperature dependence has been reported in agreement with the predictions of the spin-fluctuation theory for a 3D antiferromagnet \cite{Moriya1995}.

At low pressure $p < 1.5$~GPa the temperature of the  maximum of the resistivity $T_M$, and the N\'eel temperature $T_N$ shows only a small pressure dependence. In that pressure region the nuclear spin lattice relaxation rate $1/T_1$ is constant above $T_N$ characteristic of the response of localized moments. Above 1.5~GPa  $T_M$ starts to increase, and furthermore a decrease of $1/T_1$ is observed in the temperature range $T_N < T < T^\star$\cite{Kawasaki2002}. This indicates the formation of heavy fermion band, and the description of magnetism by the spin-fluctuation theory for itinerant electrons instead of a local moment pictures appears  to be valid. 

A small superconducting dome appears in clean samples around $p_c$. Superconductivity has been proven by resistivity ($\rho = 0$) but also by NMR experiments indicating a bulk transition. The initial slope $dH_{c2}/dT = -3.2$~T/K and the $H_{c2}(0) = 0.45$~T are also characteristic values for a superconducting state built from heavy quasiparticles with a critical temperature $T_c = 0.2$~K. Thus it has been argued that CeIn$_3$ is a prototypical example for spin fluctuation driven superconductivity at a quantum critical point.

\begin{figure}[t]
\begin{center}
\includegraphics[width=0.45\hsize,clip]{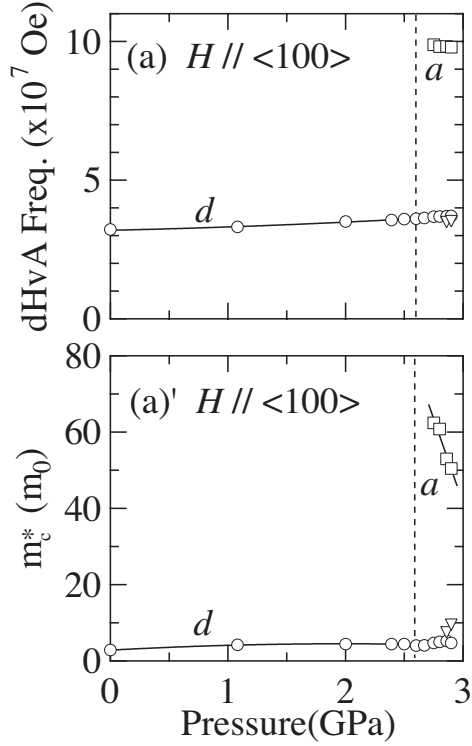}%
\caption{\label{Settai_CeIn3} Pressure dependence of the dHvA frequency (upper panel) and its cyclotron effective mass (lower panel) in CeIn3 for the magnetic field along the principal axes $H \parallel [100]$ (taken from Ref. \citealp{Settai2005}). Above the critical pressure $p_c \approx 2.6$ GPa the heavy $a$ branch could be detected. 
} 
\end{center} 
\end{figure}

Detailed studies of the Fermi surface and NQR show that the situation is more complicated. The Fermi surface of CeIn$_3$ has been investigated by  de Haas van Alphen  (dHvA) measurements up to very high magnetic fields \cite{Ebihara2004, Harrison2007, Sebastian2009} and under high pressure up to 2.8~GPa \cite{Endo2004, Settai2005}.  The experiments and comparison to band structure calculations give indications of a localized $4f$ character of the 4 $f$ electrons in the antiferromagnetic state at ambient pressure. This has also been concluded from positron annihilation radiation \cite{Biasini2003, Rusz2005}. Under high pressure above $p_c$ a change in the dHvA frequencies  has been observed which may indicate an  abrupt change in the Fermi-surface volume \cite{Endo2004, Settai2005} as shown in Fig.\ref{Settai_CeIn3}, which could be the fingerprint of a localized $\to$ delocalized transition. 

Microscopic information on the interplay of magnetism and superconductivity has been deduced from NQR studies under high pressure \cite{Kawasaki2002,Kawasaki2004,Kawasaki2008}. The transition from  the antiferromagnetic to the paramagnetic state is first order and $T_N(p_c)=1.2$~K. Phase separation into antiferromagnetism and paramagnetism appears in the vicinity of $p_c$, indicating that no second order quantum critical point appears in CeIn$_3$, but that the quantum phase transition is first order. Unconventional superconductivity occurs on both sides of $p_c$. Below $p_c$ in the coexistence regime a constant $1/T_1T$ has been reported, while above $p_c$ just below $T_c$ a power-law like $T$ dependence of $1/T_1$ indicates line nodes of the gap. 

To summarize, superconductivity appears in CeIn$_3$ in the vicinity of an antiferromagnetic to paramagnetic first order quantum phase transition. This first order transition is accompanied by an abrupt change of the Fermi surface volume at $p_c$. The  origin of the transition is the competition of antiferromagnetism and paramagnetism in the vicinity of $p_c$. However, except by NQR,  no microscopic experiments have been performed to study the coexistence range. Furthermore, there is a lack of thermodynamic investigations under high pressure which are indeed extremely challenging, due to the low $T_c = 200$~mK. 

\section{The Ce--115 family}

A breakthrough in the field of the interplay of antiferromagnetism and superconductivity has been the discovery of  pressure induced superconductivity in CeRhIn$_5$ \cite{Hegger2000} and later at ambient pressure in the CeCoIn$_5$ \cite{Petrovic2001} and CeIrIn$_5$ \cite{Petrovic2001b}. 
The crystal structure of these compounds is tetragonal and belongs to the family with the generalized stoichiometry Ce$_nM_m$In$_{3n+2m}$ in which $n$ layers of CeIn$_3$ alternate with $m$ layers of $M$In$_2$. The structure of the Ce-115 ($n=1, m=1$) family is shown in  in Fig.~\ref{crystal_structure}). Members of the double layered structure getting superconducting at ambient pressure are e.g.~Ce$_2$CoIn$_8$ ($T_c \approx 1$~K) \cite{Chen2002} or Ce$_2$PdIn$_8$ ($T_c \approx 0.70$~K) \cite{Kaczorowski2009, Uhlirova2010, Kaczorowski2010}. A common feature of all the systems is that superconductivity appears close to a magnetic quantum critical point. For spin fluctuation mediated superconductivity it has been shown that a reduction of the dimensionality favor superconductivity \cite{Monthoux2001,Monthoux2002}. This seems to be supported by the strong increase of $T_c$, from $T_c = 0.2$~K in CeIn$_3$ to $T_c \approx 1$~K in the two-layer compounds and $T_c \approx 2$~K in the Ce-115 compounds. The Fermi surface of Ce-115 compounds consists of nearly cylindrical Fermi surfaces and small  ellipsoidal ones. The cylindrical sheets reflect the two dimensional character of the electronic system being responsible for the enhancement of $T_c$ \cite{Settai2001}. It has also been noticed that an increase of the $T_c$ goes together with an increase of the $c/a$ ratio of the lattice parameters \cite{Bauer2004b}. 

Superconductivity in this structural family is not only restricted to Ce compounds, but has also been  observed in two iso-structural Pu compounds, PuCoGa$_5$ ($T_c \approx 18.5$~K) \cite{Sarrao2002} and PuRhGa$_5$ ($T_c \approx 8$~K\cite{Wastin2003}. 
In the following we concentrate on the Ce-115 compounds CeRhIn$_5$ and CeCoIn$_5$.

\section{Antiferromagnetism and Superconductivity in CeRhIn$_5$}

\subsection{High pressure phase diagram}

\begin{figure}[t]
\begin{center}
\includegraphics[width=0.6\hsize,clip]{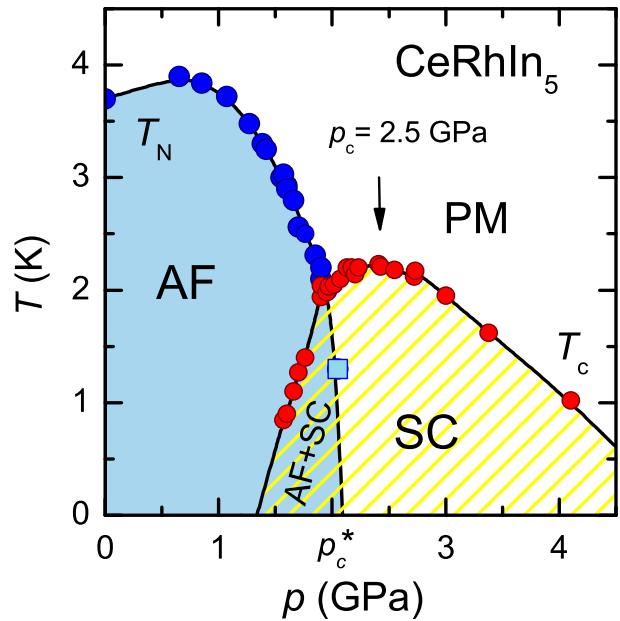}%
\caption{\label{CeRhIn5_PD} Pressure--temperature phase diagram of CeRhIn$_5$ at zero magnetic field determined from specific heat measurements with antiferromagnetic (AF, blue) and superconducting phases (SC, yellow). A coexistence phase AF+SC exists below $p_c^\star$. The blue square indicate the transition from SC to AF+SC after Ref.~\citealp{Yashima2007}.
} 
\end{center} 
\end{figure}

\begin{figure}[t]
\begin{center}
\includegraphics[width=0.6\hsize,clip]{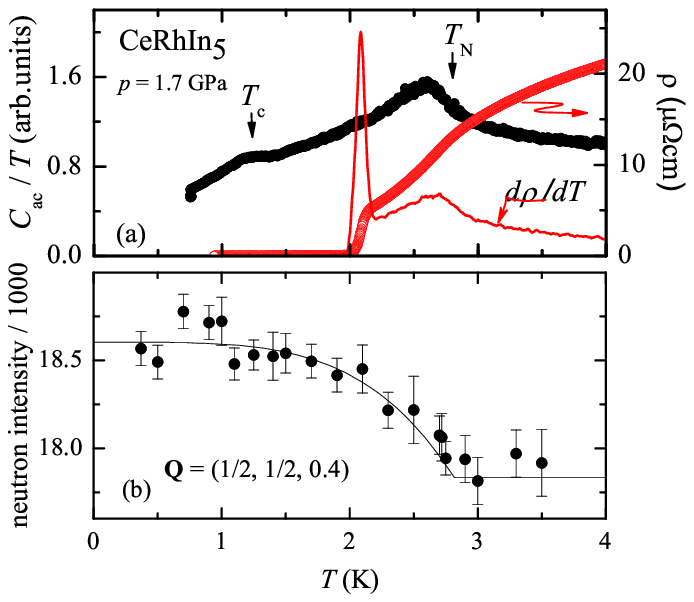}%
\caption{\label{CeRhIn5_17kbar} (a) Specific heat in $C_{ac} /T$ (left scale), resistivity (right scale) and the derivative $d\rho / dT$ of the resistivity at $p = 1.7$~GPa of CeRhIn$_5$. A tiny bulk superconducting transition in the specific heat appears at much lower temperature than the transition in the resistivity. (b) Temperature variation of the peak intensity measured for \boldmath Q \unboldmath $=(1/2, 1/2, 0.4)$ at 1.7 GPa on IN22 for a counting time of 25 min./point (see Ref.~\citealp{Raymond2008}). Lines are guide for the eyes.
} 
\end{center} 
\end{figure}

At ambient pressure CeRhIn$_5$ orders below $T_N = 3.8$~K in an incommensurate magnetic structure with an ordering vector \boldmath $Q_{ic}$\unboldmath =(0.5, 0.5, $\delta$) and $\delta=0.297$ \cite{Bao2001}. The staggered magnetic moment $\mu = 0.59\mu_B$ at 1.9~K is reduced of about 30\% in comparison to that of Ce ion in a crystal field doublet without Kondo effect \cite{Raymond2007} indicating partial Kondo screening of the moment. A moderate enhancement of the Sommerfeld coefficient of the specific heat $\gamma = 52$~mJ mol$^{-1}$K$^{-2}$) \cite{Knebel2004} and of the cyclotron masses in dHvA experiments \cite{Hall2001, Shishido2002} have been observed. The topology of the Fermi surface of CeRhIn$_5$ is similar to that of LaRhIn$_5$ and the observed Fermi volume is small indicating that the 4$f$ electrons preserve their localized character at ambient pressure in presence of weak hybridization with the conduction electrons. Modeling of the electronic structure of CeRhIn$_5$ support this view \cite{Elgazzar2004, Haule2010}.

Studying the high pressure phase diagram of CeRhIn$_5$ permits to get much deeper insights on the interplay of magnetism and superconductivity as the N\'eel and the superconducting transition temperature reach the same order of magnitude. With pressure the balance between the RKKY interaction and the Kondo interaction can be tuned and the magnetism is suppressed. In Fig.~\ref{CeRhIn5_PD} we show the pressure--temperature phase diagram obtained by ac calorimetry \cite{Knebel2006}.  The N\'eel temperature $T_N$ shows a smooth maximum around $p=0.8$~GPa and for higher pressures $T_N$ decreases monotonously. A linear extrapolation of $T_N \to 0$ would indicate a quantum critical pressure of $p_c = 2.5$~GPa in absence of superconductivity. However, pressure induced superconductivity  appears on a broad pressure range, 0.1 GPa $<p<$ 5.5~GPa.  There is also report on superconductivity at ambient pressure in very high quality samples, but the bulk nature is still under debate \cite{Chen2006,Paglione2008}. The pressure of the maximum of the superconducting transition temperature $T_c^{max} \approx 2.2$~K coincides with the linear extrapolation of $T_N \to 0$ at $p_c$. The appearance of superconductivity under pressure lead to suppress rapidly the antiferromagnetic order when $T_c = T_N \approx 2$~K at $p_c^\star \approx 2$ GPa.  Just above the vicinity $p_c^\star$ a superconducting state with $d$-wave symmetry is formed and in zero magnetic field magnetism is absent \cite{Mito2001}. This is mainly concluded from the $T^3$ temperature dependence of the nuclear spin relaxation rate $1/T_1$ and the absence of a Hebel Slichter peak. The intuitive picture is that the opening of a superconducting gap on large parts of the Fermi surface above $p_c^\star$ impedes the formation of long range magnetic order. Thus, a coexisting phase AF+SC in zero magnetic field seems only be formed if on cooling first the magnetic order is established. It should be stressed that the superconducting phase transition below $p_c^\star$ seems inhomogeneous, as large differences in $T_c$ established from different experimental probes have been reported as shown in Fig.~\ref{CeRhIn5_17kbar} for $p = 1.7$~GPa. The superconducting specific heat anomaly at $T_c$ is not at all BCS type. Thus the occurrence of homogeneous superconductivity must be associated with a modified magnetic order.

No concluding picture on the magnetic structure under high pressure has been achieved up to now from neutron scattering experiments. This may be due to very strong sensitivity of the ground state to non-hydrostatic pressure conditions \cite{Raymond2008, Aso2009}. The magnetic ordering vector changes with pressure and \boldmath $Q_{ic}$\unboldmath =(0.5, 0.5, $\delta=0.4$)at $p = 1.7$~GPa   and the ordered moment $\mu < 0.2\mu_B$ is significantly reduced (see Fig.~\ref{CeRhIn5_17kbar})\cite{Raymond2008}. In Ref.~\citealp{Aso2009} it has been stated that when the temperature is
lowered at $p\approx 1.5$ GPa, the magnetic Bragg peak with $\delta = 0.326$ rapidly disappears and instead a new peak with $\delta = 0.39$ suddenly emerges at a temperature close to the superconducting transition temperature. This points to a direct interplay of superconductivity and magnetism. However, neutron diffraction experiments failed up to now to get closer to the first critical pressure $p_c^\star$. 

Detailed NQR experiments enlightened the AF+SC state \cite{Yashima2007, Yashima2009} and give strong evidences for  homogeneous coexistence of superconductivity and antiferromagnetic order. Here is has been nicely shown that the magnetic structure gets commensurate $(\delta = 0.5)$ when superconductivity appears for $p > 1.7$ GPa \cite{Yashima2009}. Below $p_c^\star$ it is observed that the phase transition on cooling  from AF to AF+SC looks quite inhomogeneous, but far below $T_c$, the spin-lattice relaxation $1/T_1$ is homogeneous independent on the local site what is also a nice hint of the coexistence of both states below $p_c^\star$ \cite{Kawasaki2003,Yashima2007,Yashima2009}. By analyzing in detail the NQR spectra it could be shown that even at $p=0$ a small commensurately ordered volume fraction exists inside the incommensurate order which might be responsible for the observed superconducting transitions at $p=0$ \cite{Chen2006,Paglione2008}. The coexistence phase seems strongly coupled to the commensurate magnetic order. Thus in the phase diagram of CeRhIn$_5$ a tetra-critical point appears \cite{Yashima2007} and $p_c^\star$ is the pressure where the magnetic order is rapidly suppressed. 

\begin{figure}[t]
\begin{center}
\includegraphics[width=0.42\hsize,clip]{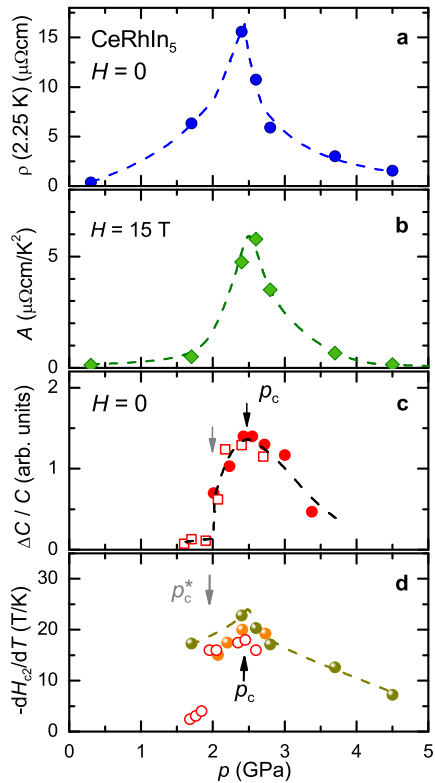}%
\caption{\label{CeRhIn5_parameter} Pressure dependence of (a) the resistivity at $T= 2.25$~K just above the superconducting transition, (b) the $A$ coefficient of the resistivity measured at a field $H=15$T far above the upper critical field $H_{c2}$. 
(c) Specific heat jump $\Delta C/C$ at the superconducting transition as function of pressure. (Different symbols corresponds to different experiments.) (d) Pressure dependence of the initial slope of the upper critical field at the superconducting transition. Full circles are from the resistivity and specific heat experiments \cite{Knebel2008, Knebel2006},  open circles are taken from Ref.~\citealp{Park2008b}, the pressure of Ref.~\citealp{Park2008b} has been normalized to our experiments.
} 
\end{center} 
\end{figure}

Experimental evidence for the quantum critical pressure $p_c$ (defined from the extrapolation of $T_N \to 0$) can be obtained from the normal state, but also from the superconducting properties. A first indication of $p_c$ is the strong enhancement of the resistivity just above the superconducting transition (see Fig.~\ref{CeRhIn5_parameter}a) due to the enhancement of the scattering caused by critical fluctuations \cite{Miyake2002}. Above the superconducting transition the resistivity shows an unusual sub-linear temperature dependence \cite{Knebel2008, Park2008d} which is an indication of critical valence fluctuation \cite{Watanabe2010a} and has also been discussed in terms of unconventional quantum criticality \cite{Park2008d}. Another striking feature is the strong enhancement of the inelastic scattering term in the resistivity as shown in Fig.~\ref{CeRhIn5_parameter}b) where the pressure dependence of the $A$ coefficient of the resistivity $\rho = \rho_0 + AT^2$ obtained at high magnetic field $H = 10$~T is plotted. Again, a strong maximum appears at $p_c$, and below $p_c$ the average effective mass derived from $\sqrt{A}/m^\star = const$ increases similarly to the
effective mass $\beta_2$ branch obtained by dHvA measurement \cite{Knebel2008} indicating strongly that the enhancement of the $A$ coefficient is coupled to a band effect \cite{Watanabe2010}. 

Figure \ref{CeRhIn5_parameter}c-d) shows the pressure dependence of the specific heat anomaly $\Delta C/C$ at $T_c$ in zero magnetic field and of the initial slope of the superconducting upper critical field . Two points are remarkable: firstly both quantities show pronounced peaks at $p_c$ indicating a maximum of the effective mass. Secondly, there is a strong increase of both quantities just at the critical pressure $p_c^\star$ where the magnetic order collapses. In Ref.~\citealp{Park2009} an analysis of the entropy at the magnetic and the superconducting transition as function of pressure shows that in the same way as the  entropy at $T_N$ decreases the entropy at $T_c$ increases. This clearly shows the close coupling of both orders, and explains the strong increase of the specific heat jump at $p_c^\star$. The initial slope of the upper critical field $H_{\rm c2}'=(dH_{\rm c2}/dT)_{T=T_{\rm c}} \approx T_c/v_{\rm F}^2$ is a good measure of the average Fermi velocity $v_F$ in a plane perpendicular to the applied field and thus a strong change of $v_F$ occurs at $p_c^\star$, as $T_c$ varies rather smoothly in that pressure range (see Fig.~\ref{CeRhIn5_PD}).  This shows that a strong change in the electronic structure already appears at this pressure in zero magnetic field. 

\subsection{Magnetic field effect: Re-entrance of the antiferromagnetic phase}

\begin{figure}[t]
\begin{center}
\includegraphics[width=1\hsize,clip]{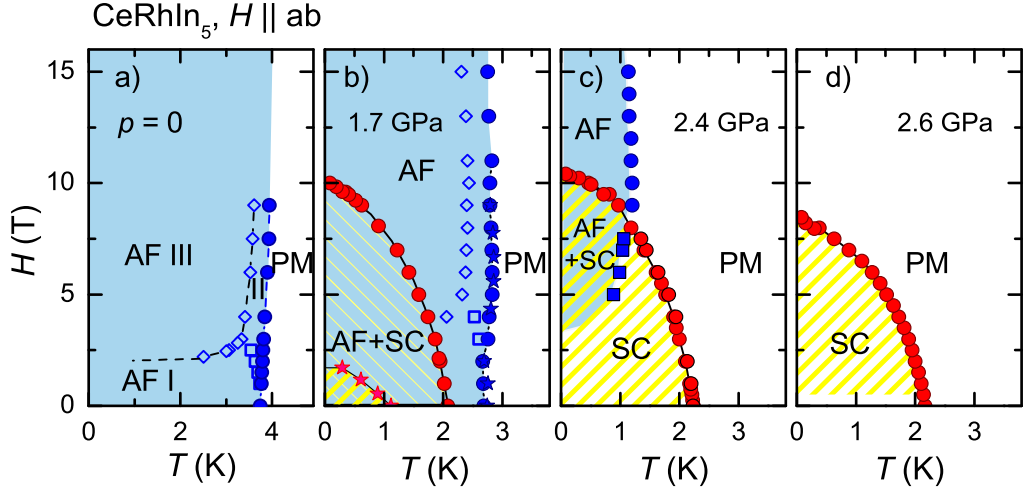}%
\caption{\label{CeRhIn5_field} Magnetic field phase diagram of CeRhIn$_5$ at different pressures for a magnetic field applied in the $ab$ plane (blue symbols for magnetic transitions, red symbols for superconductivity). a-b) The magnetic phase diagram is almost unchanged compared to $p=0$ up to $p_c^\star$. However, $H_{c2} (T)$ detected by specific heat (red stars, taken from Ref.\citealp{Park2008b}) is much lower that detected by resistivity. Remarkably magnetic order is induced inside the superconducting dome in the pressures $p_c^\star < p < p_c$, as shown for $p=2.4$~GPa in c). d) For $p> p_c$ no magnetic order appears.
} 
\end{center} 
\end{figure}

\begin{figure}[t]
\begin{center}
\includegraphics[width=0.6\hsize,clip]{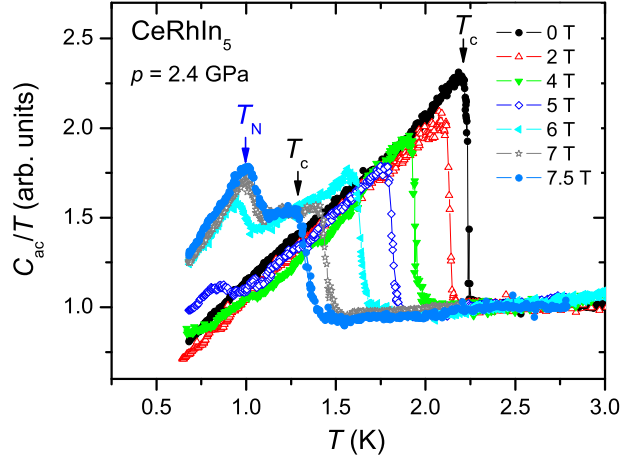}%
\caption{\label{CeRhIn5_Cac24} Specific heat divided by temperature of CeRhIn$_5$ for different magnetic fields applied in the basal plane. For $H > 4$~T a new phase transition appears inside the superconducting phase. 
} 
\end{center} 
\end{figure}

Figure \ref{CeRhIn5_field} shows the magnetic field phase diagram of CeRhIn$_5$ for different pressures. At ambient pressure the magnetic properties have been studied in detail up to 60 T \cite{Takeuchi2001}. For magnetic field $H \perp c$ the critical field to reach the field induced paramagnetic state is $H_m \approx 52$ T. For $H \parallel c$ $H_m$ has not been determined. The magnetic phase diagram for $H\perp c$ shows three different phases and the magnetic structures have been determined by neutron scattering \cite{Raymond2007}. At low pressure the incommensurate structure (AF I) changes to a commensurate (AF III) with \boldmath $Q_{c}$\unboldmath =(0.5, 0.5, 0.25) at a field $H_{ic} \approx 2.25$~T. Phase AF II shows also the incommensurate ordering vector, but a collinear sine wave is formed. 

As discussed already above, under application of pressure the magnetic phase diagram does not change up to $p=1.7$ GPa, and three distinct phases do appear under magnetic field. However, the magnetic structures have not been determined up to now. Figure \ref{CeRhIn5_field}b) demonstrates that the occurrence of superconductivity, and the determination of the magnetic phase diagram is very sensitive to the pressure condition and to the experimental probe. Whereas in specific heat experiment only a small superconducting region appears (see Fig.~\ref{CeRhIn5_17kbar}), in resistivity experiments the superconducting region in the $(T,H)$ plane is much larger:  the superconducting phase diagram determined by resistivity is only comparable to the bulk superconducting state for $p>p_c^\star$. 
A spectacular observation is the re-entrance of magnetism in the pressure range $p_c^\star < p < p^\star$ (see Fig.~\ref{CeRhIn5_field}c)~\cite{Park2006,Knebel2006}. At zero field the ground state is a superconductor as shown in Fig.~\ref{CeRhIn5_Cac24} for $p = 2.4$~GPa, but for $H>4$~T a second sharp anomaly is observed inside the superconducting state. This field induced, probably magnetic, state seems to coexist peacefully with superconductivity in that pressure range, and extends above the upper critical field \cite{Knebel2008}. Above the critical pressure $p_c$, no field induced state has been observed. This lead to the proposal that the field induced state is strongly connected to the the vortex state and to the Fermi surface properties, and also points the relative weakness of superconductivity under magnetic field in comparison to the robustness of antiferromagnetism, even up to the critical pressure $p_c$ (weak field dependence of $T_N(H)$ by comparison to $T_c(H)$). Furthermore, we want to mention that in CeRhIn$_5$  field induced phases appear for both crystallographic directions, by contrast to the case of CeCoIn$_5$ which will be discussed below. 

\begin{figure}[t]
\begin{center}
\includegraphics[width=0.6\hsize,clip]{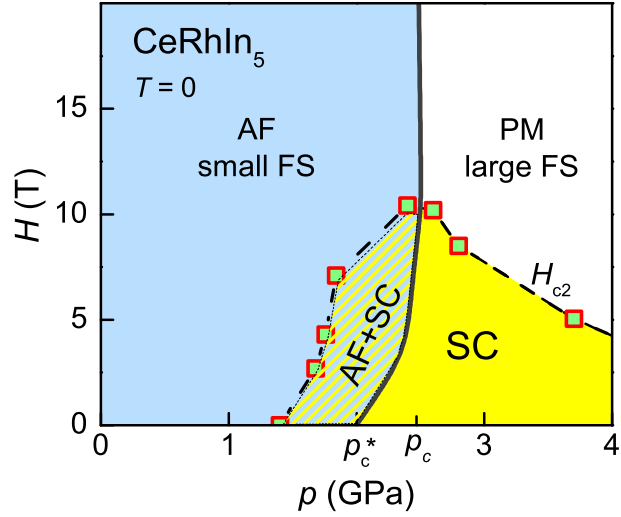}%
\caption{\label{CeRhIn5_FS} ($H, T$) phase diagram of CeRhIn$_5$ at $T=0$ indicating the Fermi surface topology in the different states of the phase diagram. The boundary between the  ''small`` Fermi surface (localized description of the 4$f$ electron) and of the itinerant paramagnetic  phase (''large`` Fermi surface with itinerant description of the 4$f$ electron) is indicated by thick black line. One yet unsolved question is the Fermi surface topology in the AF+SC state with the strong interplay between antiferromagnetism and superconductivity. One can speculate that at $H = 0$ an itinerant Fermi surface persists down to $p_c^\star$ as indicated. The different magnetic structures are not indicated. 
} 
\end{center} 
\end{figure}

An intuitive picture of the coexistence regime under magnetic field is that magnetism nucleates in the vortex cores and long range magnetic order would occur spontaneously at high enough magnetic field where vortex cores almost start to overlap. A theoretical description of the coexistence regimes in CeRhIn$_5$ may be given in the framework of SO(5) theory (see Ref.~\citealp{Demler2004} and references therein). In this model, developed for high $T_c$ superconductors, the spin and charge modulations of the system are interpreted as textures of the SO(5) superspin as it rotates in SO(5) space. Thus superconducting (the gap $\Delta_{SC}$) and magnetic order parameter (staggered magnetization $M_q$) are coupled so that $\mid \!\! \Delta_{SC}\!\! \mid ^2 + \mid \!\! M_q \!\! \mid ^2 = 1$. However, detailed predictions for microscopic experiments to test the theory are still missing. 

The re-entrance of the magnetism in the superconducting phase is also strongly coupled to Fermi-surface properties. By dHvA experiments under high magnetic field it has been shown that at $p_c$ the volume of the Fermi surface changes abruptly \cite{Shishido2005} from a small, $f$ localized Fermi surface to large, $f$ itinerant, one and that the volume of the Fermi surface above $p_c$ is comparable to that of CeCoIn$_5$. The dHvA frequencies change abruptly at the critical pressure $p_c$ and the cyclotron mass of the heavy $\beta_2$ and the $\alpha_{2,3}$ branches are strongly enhanced on approaching $p_c$. Above $p_c$ only the $\alpha$ branch has been observed. Other branches are not detected mainly due to the existence of a large cyclotron effective mass close to 100 $m_0$. Figure~\ref{CeRhIn5_FS} gives a sketch of the $(H,T)$ phase diagram at $T=0$. A still open problem is the Fermi surface in the coexistence regime AF+SC. Furthermore, inside the AF phase new additional frequencies have been observed for $p>1.7$~GPa which may account for changes in the magnetic structure. However, we want to emphasize that dHvA experiments are performed for fields $H>H_{c2}$ and thus the abrupt change of the Fermi surface at zero field may already appear at $p_c^\star$ as indicated in Fig.~\ref{CeRhIn5_FS}. A strong indication for this perception is that strong variations in the electronic states appear already at $p_c^\star$ (suppression of magnetism and accompanied formation of a superconducting ground state). Future experiments have to clarify this point. Finally, we want to mention that above the critical pressure $p_c$ no magnetic phase has been observed indicating that magnetism is connected to the ''small`` Fermi surface and the magnetic phase diagram collapses at $p_c$.


\subsection{Concluding remarks on CeRhIn$_5$}
High pressure studies on CeRhIn$_5$ allowed very deep insights in the interplay between antiferromagnetism and superconductivity due to the fact that their ordering temperatures becomes very close in the critical pressure region and also that the critical pressure can be reached in large volume pressure cell.  This allows detailed ac calorimety, quantum oscillation, and NQR experiments to draw precisely the phase boundaries and the correlations with Fermi surface or or magnetic structures changes. The broadly accepted picture is that antiferromagnetism and superconductivity homogeneously coexist below $p_c^\star$ when the magnetic structure is commensurate with the crystal structure. When the superconducting transition temperature $T_c$ as function of pressure is higher than the N\'eel temperature, magnetism is rapidly suppressed at this critical pressure $p_c^\star$ and the ground state is a $d$ wave superconductor with line nodes. Thus, the collapse of the antiferromagnetic phase is closely related to the pressure induced superconducting state. The maximum of $T_c$ coincides with a maximum of the effective mass of the charge carriers, and under high magnetic field a change of the Fermi surface occurs at this pressure. In the pressure region $p_c^\star < p < p_c$ long range magnetic order can be recovered by the application of high magnetic field even inside the superconducting state, but the antiferromagnetic order also persists for fields far above the upper critical field $H_{c2}$. This induced phase disappears due to a reconstruction of the Fermi surface at the expected magnetic quantum critical point in absence of superconductivity.

\begin{figure*}[tb]
\begin{center}
\begin{minipage}{0.49\hsize}
\begin{center}
\includegraphics[width=.9\hsize,clip]{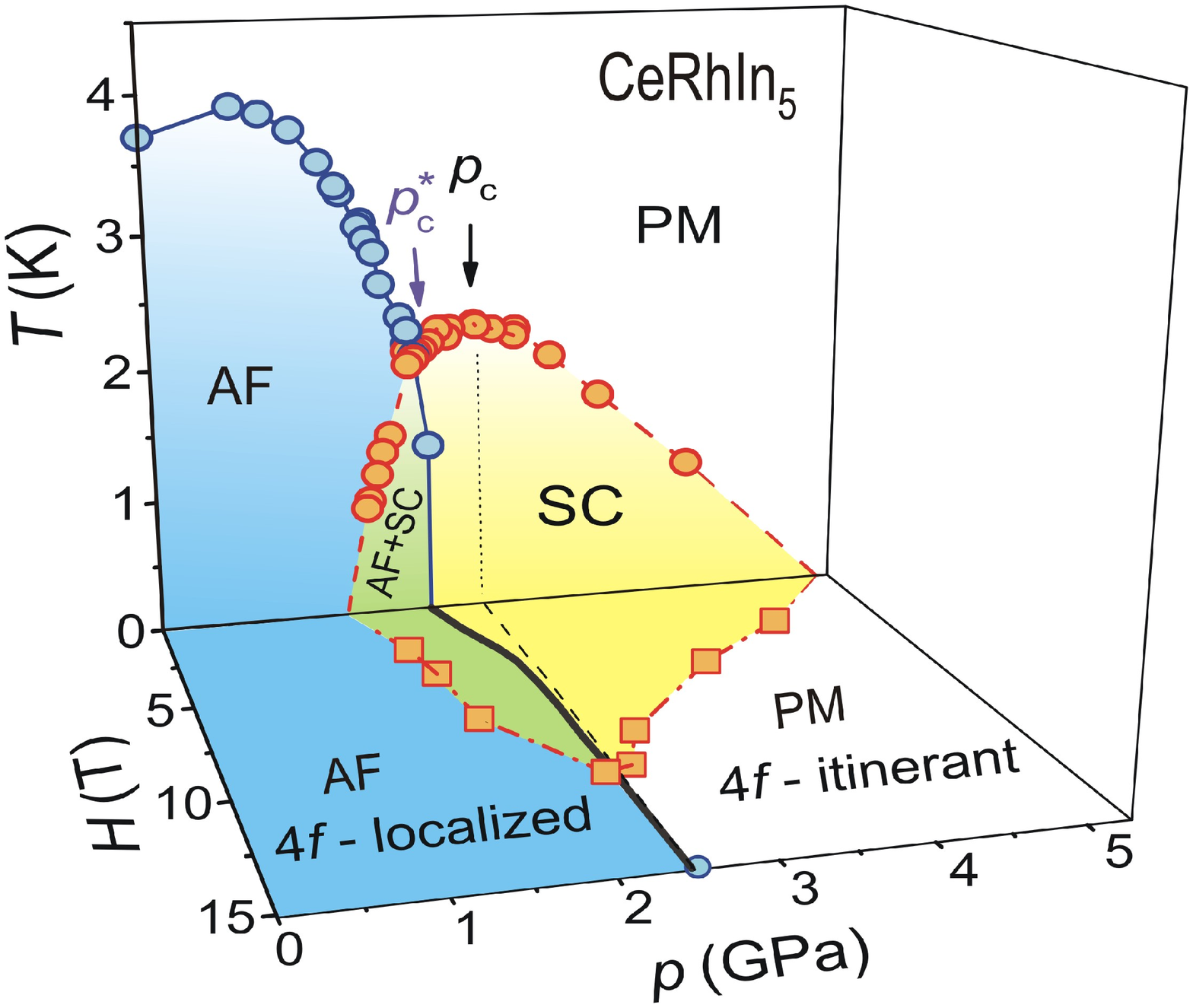}
\end{center} 
\end{minipage}
\begin{minipage}{0.49\hsize}
\begin{center}
\includegraphics[width=.8\hsize,clip]{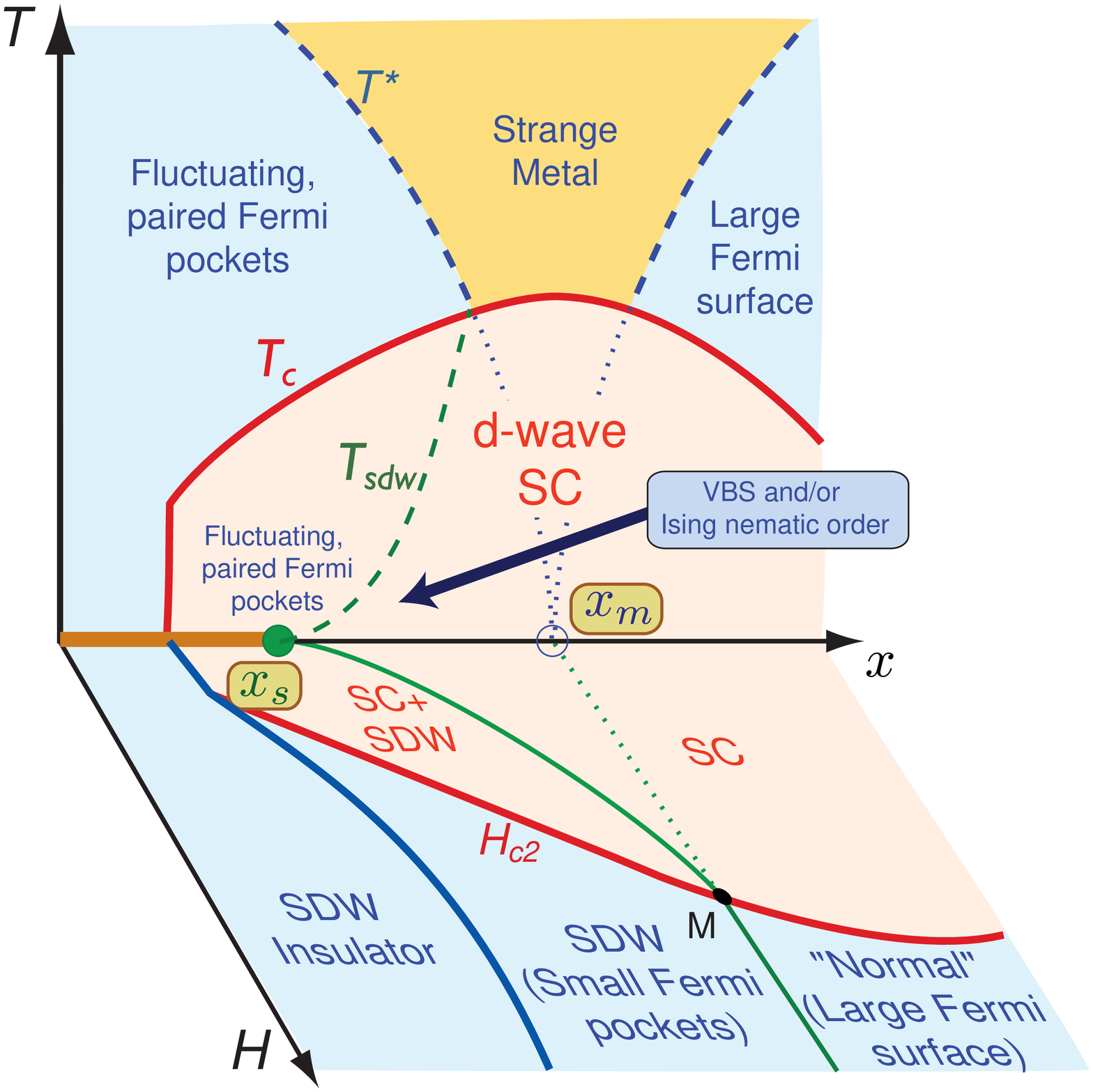}
\end{center} 
\end{minipage}
\caption{\label{Sachdev}(Left panel) Combined temperature, pressure and field $H\perp c$ phase diagram of CeRhIn$_5$  with antiferromagnetic (blue), superconducting (yellow), and coexistence AF+SC (green) phases. The thick black line in the $H-p$ plane indicates the proposed line where the Fermi surface changes from 4$f$ "localized" (small Fermi surface and topology comparable to LaRhIn$_5$), to 4$f$ "itinerant" (large Fermi surface as in CeCoIn$_5$).  (Right panel) Proposed phase diagram of the high $T_c$ cuprates showing the interplay between superconductivity (SC), spin density order (SDW), and Fermi surface configuration as function of carrier density ($x$), temperature ($T$), and magnetic field $(H)$ perpendicular to the CuO$_2$ layers (taken from S. Sachdev, Ref.~\citealp{Sachdev2009}). }
\end{center} 
\end{figure*}

We want to compare the observed phase diagram of CeRhIn$_5$ to that of other classes of strongly correlated electron systems. The generic phase diagram of the iron based pnictides with ThCr$_2$Si$_2$ structure, e.g.~Ba(Fe$_{1-x}$Co$_x$)$_2$As$_2$, is very similar to that of CeRhIn$_5$ (see e.~g.~Refs.\cite{Ni2008,Chu2009}). The parent compound of this family of pnictides, BaFe$_2$As$_2$, exhibits antiferromagnetic spin density wave order below $T_N = 140$~K which coincides with a structural transition to an orthorhombic low temperature structure. Under doping or high pressure the structural transition is detached from the magnetic transition and in both cases superconductivity is induced. The maximum of $T_c$ coincides with the onset of a purely superconducting ground state at optimal doping. The coupling  of antiferromagnetism and superconductivity has been shown by neutron diffraction experiments \cite{Fernandez2010} with a strong reduction of the intensity of the magnetic Bragg peak below $T_c$. The superconducting order parameter in pnictides is most probably a two-band, presumably sign changing $s$-wave function (see e.g.~recent reviews \cite{Paglione2010, Mazin2010}. However, many details of the phase diagrams of pnictides have still to be enlighten, such as e.g.~the Fermi surface properties in the different phases or the role of the magneto-elastic coupling.  Progress will also depend on the possibility to grow high quality single crystals. 

Finally we want to compare the phase diagram of CeRhIn$_5$ with one proposed for the high $T_c$ cuprates. A striking point is the similarity between the proposed phase diagrams of the two systems which are shown in Fig.~\ref
{Sachdev}. (A significant difference is of course that the parent compound of cuprates is an antiferromagnetic  Mott insulator.) The pressure variable in CeRhIn$_5$ is replaced by the carrier concentration for the cuprates. In zero magnetic field the magnetic quantum critical point at $p_c$ or $x_m$ for the cuprates is hidden by the onset of superconductivity at high temperatures. The underlying quantum critical point determines the normal state properties, the observed non Fermi liquid behavior with the linear $T$ dependence of the electrical resistivity in the "strange" metallic regime. The formation of a superconducting ground state has also influence on the spin density wave order and shifts its disappearance to a concentration $x_s <x_m$ \cite{Moon2009} comparable to the pressure shift of disappearance of magnetism in CeRhIn$_5$ from $p_c$ to $p_c^\star$ in zero magnetic field. 

In both classes of systems at a magnetic fields higher than the upper critical field $H_{c2}$ a Fermi surface instability is connected to the collapse of the magnetic phase. The case of CeRhIn$_5$ has been discussed above as a transition from a $4f$ localized to a $4f$ itinerant state with a large Fermi surface. In the high $T_c$ cuprates a change of the Fermi surface at high field through $x_m$ appears from small Fermi surface pockets in the underdoped regime $x< x_m$ to a large Fermi surface in the overdoped regime \cite{Hussey2003,Vignolle2008,Doiron-Leyraud2007,Yelland2008,Bangura2008,Sebastian2008}. 

Under magnetic field the appearance of vortices favors the re-entrance of magnetism in the superconducting state: for CeRhIn$_5$ this may happen in the pressure range $p_c^\star < p <p _c$, for different high $T_c$ materials this has been experimentally shown by neutron scattering experiments \cite{Lake2001, Lake2002, Kang2003, Chang2009}, NMR experiments \cite{Mitrovic2001, Kakuyanagi2003, Mitrovic2003}, or muon spin  resonance ($\mu$SR) \cite{Miller2002, Miller2006}. The boundary between SC+SDW and SC phase moves from $x_s$ to $x_m$ (M-point) when $H$ reaches $H_{c2} (x_m)$. Thus the line between the points $(x_s, H=0), (x_m, H_M)$ in the $x, H$ plane at $T =0$ between SDW+SC and the pure superconducting phases delimits two different regimes: small Fermi surface pockets (left of the line) in the SDW+SC phase and a large Fermi surface in the sole SC phase. As discussed above, a similar scenario may be valid for CeRhIn$_5$.

\section{Unconventional Superconductivity in CeCoIn$_5$}
\label{UCS_CeCoIn5}
\begin{figure}[t]
\begin{center}
\includegraphics[width=0.6\hsize,clip]{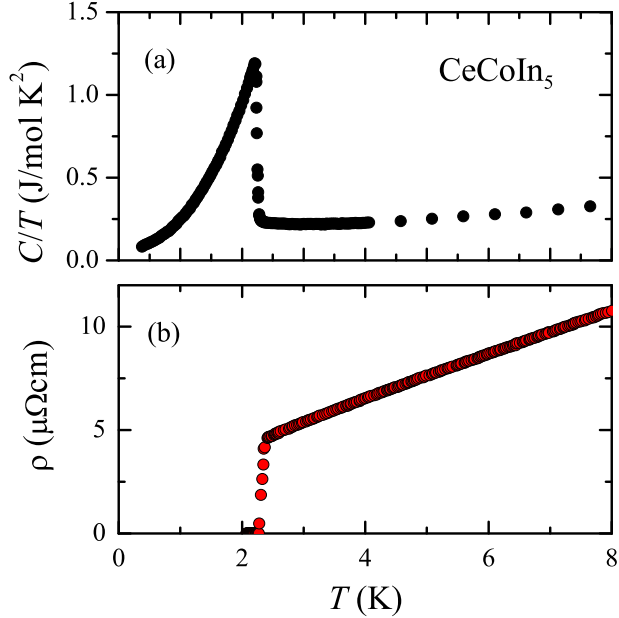}%
\caption{\label{CeCoIn5_C_rho} (a) Specific heat divided by temperature of CeCoIn$_5$. (b) Temperature dependence of the resistivity $\rho$ vs. $T$ of CeCoIn$_5$. Above the superconducting transition $\rho$ shows a linear temperature dependence up to $T \approx 10$~K.
} 
\end{center} 
\end{figure}

Soon after the discovery of pressure-induced superconductivity in CeRhIn$_5$, the Los Alamos group discovered that CeCoIn$_5$ is superconducting already at ambient pressure. The superconducting transition temperature $T_c = 2.3$~K is yet the highest for all Ce based heavy fermion superconductors \cite{Petrovic2001}. Compared to CeRhIn$_5$ the lattice volume is smaller, which explains the stronger hybridization. This more general view has also been confirmed by electronic structure calculations with density-functional theory combined with dynamical mean-field theory (DFT+DMFT) for all three Ce-115 compounds \cite{Haule2010}. These calculations clearly show that the Rh compound is most localized while CeIrIn$_5$ is most itinerant but this is not only an effect of the lattice structure, but also a consequence of the the chemistry of the transition metal ion (mainly the difference between 3$d$, 4$d$ and 5$d$ orbitals). 

The normal state properties of CeCoIn$_5$ indicate strong deviations from a Fermi-liquid behavior and the closeness of the system to an antiferromagnetic QCP. The electrical resistivity shows almost a linear temperature dependence above the SC transition up to about $T=10$~K \cite{Petrovic2001,Sidorov2002} (see Fig.~\ref{CeCoIn5_C_rho}). The specific heat coefficient just above the superconducting transition temperature is enhanced $C/T = 290$~mJ/(mol K$^2$) but no coherent Fermi liquid is realized when superconductivity appears. It is the Cooper pairing which will drive to the coherence as shown in the huge jump of the superconducting transition in zero field at $T_c$, $\Delta (C/T)/(C/T)\!\!\mid _{T_c} = 4.5$, far above the weak coupling limit predicted for  if $C$ is equal to $\gamma T$. 
Under magnetic field just above the upper critical field $C/T$ increases with $C/T \propto - \ln T$ and $\gamma \gg 1000$~mJ/(mol K$^2$) is reached when superconductivity is suppressed for both crystallographic directions \cite{Petrovic2001, Bianchi2003c, Ronning2005}. The maximum value of $\gamma (H)$ appears right at the upper critical field $H_{c2} (0)$ \cite{Paulsen2011}. 

The Fermi surface of CeCoIn$_5$ has been studied in detail by quantum oscillation \cite{Settai2001, Hall2001, McCollam2005}. It shows a nearly cylindrical and, therefore, quasi-two-dimensional sheet and several three dimension sheets.  There is very good agreement with the experimentally observed Fermi surface and band-structure calculations treating the $4f$ electrons as itinerant states  \cite{Settai2001, Elgazzar2004}. Such a heavy $4f$ band has been observed by angle-resolved photoemission spectroscopy (ARPES) \cite{Koitzsch2009}. 

\begin{figure*}[tb]
\begin{center}
\begin{minipage}{0.5\hsize}
\begin{center}
\includegraphics[width=1\hsize,clip]{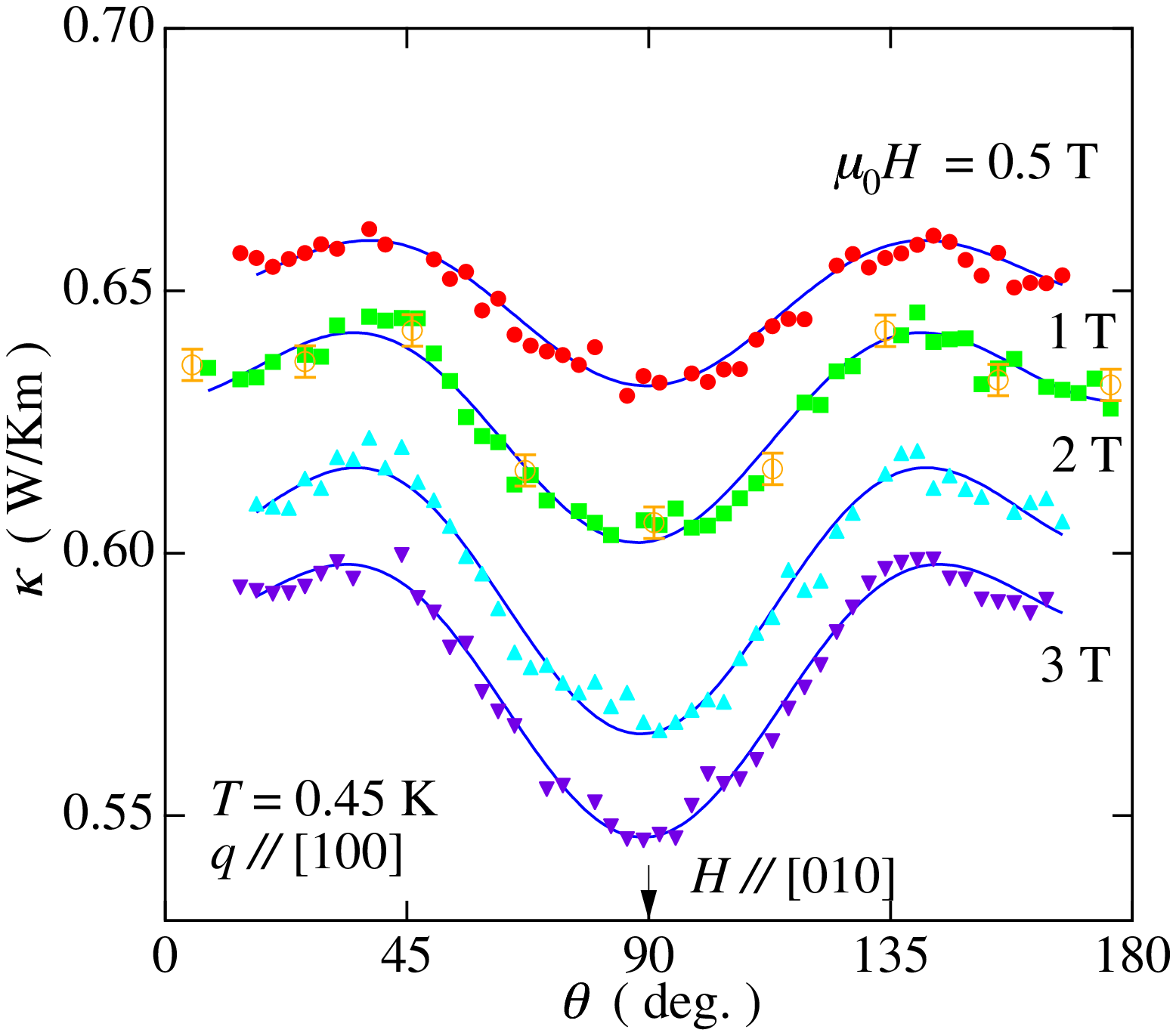}
\end{center} 
\end{minipage}
\begin{minipage}{0.4\hsize}
\begin{center}
\includegraphics[width=.5\hsize,clip]{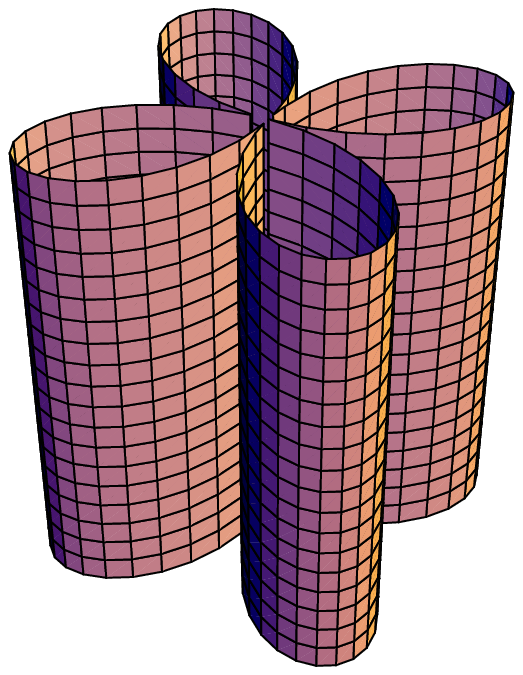}
\end{center} 
\end{minipage}
\caption{\label{Koichi}(Left panel)In plane angular dependence of the thermal conductivity $\kappa (H, \theta)$ for CeCoIn$_5$ at $T = 0.45$~K indicating the four fould symmetry (taken from Ref.~\citealp{Izawa2001}). (Right panel) Schematic presentation of the superconducting $d_x^2-d_y^2$ order parameter with vertical line nodes.  }
\end{center} 
\end{figure*}

Superconductivity in CeCoIn$_5$ has been shown to be unconventional with $d_{x^2-y^2}$ symmetry (see Fig.~\ref{Koichi}). This has been concluded from the observation of power law dependences of the specific heat and thermal conductivity \cite{Movshovich2001}, from the fourfold symmetry of the angular dependence  of specific heat in the basal plane \cite{Aoki2004, Sakakibara2007} and thermal conductivity \cite{Izawa2001}, and from the detection of a sharp resonance in neutron scattering experiments which appears in the superconducting phase \cite{Stock2008}. Microscopic evidence for the $d$-wave state is given by the absence of a Hebel Slichter peak and the $1/T_1 \propto T^3$  temperature dependence in the NMR relaxation rate \cite{Kohori2001}. The large initial slope of the upper critical field $H_{c2}$ indicates high effective masses characteristic for  the heavy fermion superconductivity with short coherence length $\xi_a = 82$~\AA\ and $\xi_c = 32$~\AA\ \cite{Settai2001}. The anisotropy of the upper critical field can be explained by an effective mass model.  The mean free path $\ell > 1000$~\AA\ shows that superconductivity is in the clean limit. Details of the upper critical field and the superconducting phase diagram will be discussed below. 

Microscopic indications for the closeness to an antiferromagnetic QCP and the strong antiferromagnetic spin fluctuations come from the temperature dependence of the spin-lattice relaxation rate $T_1$ which varies over a large temperature range from $T_c$ up to 100 K as $1/T_1 \propto T^{1/4}$ \cite{Kohori2001, Kawasaki2003b}. Systematic  NMR experiments have shown  that antiferromagnetic spin fluctuations play an active role in the superconducting pairing \cite{Curro2003}. It has been shown that a strong correlation exists between the superconducting pairing symmetry and the magnetic anisotropy in $f$ electron systems. For the Ce-115 family the favorable magnetic anisotropy is antiferromagnetic XY-type which favors $d$-wave singlet superconductivity \cite{Kambe2007, Sakai2010, Kambe2011}. Further microscopic evidence that a Fermi-liquid state is not reached even at low temperatures have been concluded from dHvA experiments where strong deviations from the usual Fermi-liquid behavior of the amplitude of the dHvA oscillations have been reported \cite{McCollam2005}. These experiments indicate that CeCoIn$_5$ is very close to an antiferromagnetically ordered state, and it has been shown that small doping by Cd or Hg on the In site induced magnetic order \cite{Pham2006, Bauer2008}. The intuitive coupling of magnetism and superconductivity in CeCoIn$_5$ has been nicely demonstrated by neutron spectroscopy with the appearance of a sharp inelastic peak that appears for an energy of 0.6 meV ($\approx 3k_BT_c$) at the antiferromagnetic position \boldmath Q \unboldmath ~$=(1/2, 1/2, 1/2)$ \cite{Stock2008, Panarin2009}. 

\subsection{Superconducting state under magnetic field}

\begin{figure}[t]
\begin{center}
\includegraphics[width=0.6\hsize,clip]{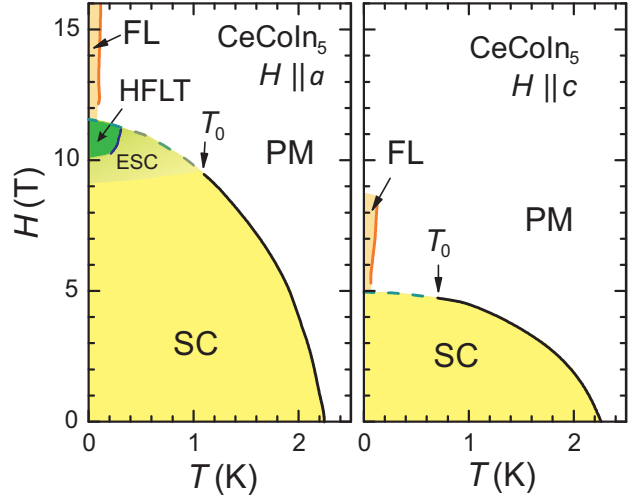}%
\caption{\label{CeCoIn5_Hc2} $H-T$ phase diagram of CeCoIn$_5$ for magnetic field $H \parallel a$ and $H \parallel c$ axis. For fields in both directions the superconducting transition is strongly Pauli limited and a first order superconducting transition has been observed below $T_0$ (dashed line). Remarkably, for $H \parallel a$ a new HFLT phase, has been observed confined to the superconducting phase has been observed. From neutron diffraction and NMR experiments incommensurate magnetic ordering is confirmed in the HFLT phase \cite{Kenzelmann2008, Kenzelmann2010, Young2007, Koutroulakis2010}. ESC indicates the exotic superconducting crossover regime \cite{Koutroulakis2010} where already strong antiferromagnetic fluctuations appear.  Above the upper critical field $H_{c2}(0)$ for both directions Fermi-liquid behavior has been observed above $H_{c2}(0)$ \cite{Paglione2003, Bianchi2003c, Kim2001, Ronning2005,Paglione2006,Howald2011}. 
} 
\end{center} 
\end{figure}

In most superconductors, the upper critical field $H_{c2}$ to suppress superconductivity is largely dominated by the orbital pair breaking due to the large Fermi velocities. The Zeeman energy of the electron spin in a magnetic field has also influence on the upper critical field: when the Zeeman energy of the electrons in the normal state gets larger than the superconducting condensate energy [the so-called Pauli limitation is reached, leading to a superconducting to normal state transition]. The relative strength of the Pauli and orbital limiting fields is reflected by the Maki parameter $\alpha = \sqrt{2} H^{orb}_{c2} / H^P_{c2}$. If the orbital limit is totally quenched, as e.g. in two-dimensional thin films, $H_{c2}$ is expected become first order below $T_0 = 0.56 T_c$. The large Zeeman splitting can also give rise to the formation of a peculiar spatially modulated SC state, as predicted by Fulde-Ferrell-Larkin-Ovschinikov (FFLO) \cite{Fulde1964, Larkin1964}. Basically the magnetic polarization will induce an instability of the Cooper pairs which have to be formed from spin up and spin down components, inducing a FFLO modulated phase along the direction of the external field. 

The $(H-T)$ phase diagram of CeCoIn$_5$ shows several peculiarities manifesting the strong interplay of magnetism and superconductivity. Figure~\ref{CeCoIn5_Hc2} shows the phase diagrams for $H \parallel a$ and $H\parallel c$. Indeed, the superconducting transition gets first order for $T < T_0 = 0.7$~K and $T < T_0 = 1.1$~K for $H\parallel c$ and $H \perp c$, respectively, indicating the strong Pauli limit \cite{Tayama2002, Bianchi2002, Bianchi2003c}. The strong Pauli paramagnetic pair breaking is essential to understand the anomalous field dependence of the vortex lattice form factor \cite{Bianchi2008, Michal2010, Ikeda2010}. It will drive a Pauli depairing of the Cooper pair in a first order metamagnetic transition. 

\begin{figure}[t]
\begin{center}
\includegraphics[width=0.6\hsize,clip]{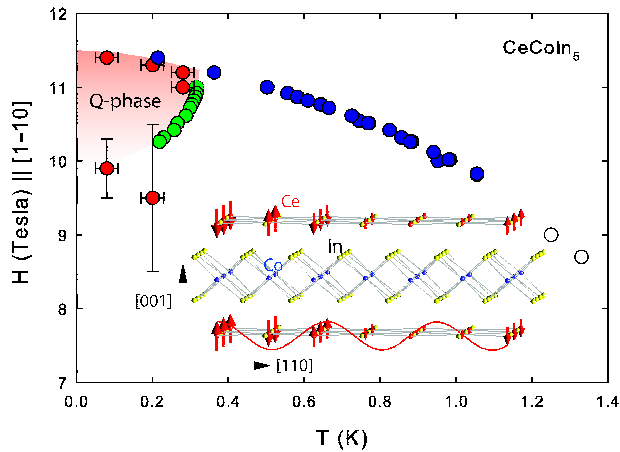}%
\caption{\label{Kenzelmann} Enlaged view on th $H-T$ phase diagram close to $H_{c2}(0)$ of CeCoIn$_5$ for magnetic field $H \parallel a$. (Inset) Magnetic structure of CeCoIn5 at $T = 60$~mK and $H =
11$~T. The red arrows show the direction of the magnetic moments, solid red line indicates the modulation of the amplitude of the  magnetic moment along the $c$-axis. Yellow and blue circles
indicate the position of the In and Co ions. (Figure is taken from Ref.~\citealp{Kenzelmann2008}.)
} 
\end{center} 
\end{figure}

A new phase appears for $H$ in the $a,b$ plane at high magnetic fields and low temperatures (HFLT)  inside the superconducting phase and this phase does not extend above $H_{c2}$ by contrast to the case of CeRhIn$_5$.  Therefore the HFLT phase (named $Q$-phase in Ref.~\citealp{Kenzelmann2008}) has first been interpreted as the realization of a FFLO state and caused intense research in the field. The HFLT phase has been identified by different experimental probes including specific heat \cite{Radovan2003, Bianchi2003c}, thermal conductivity \cite{Capan2004}, ultrasound velocity \cite{Watanabe2004}, and NMR \cite{Kakuyanagi2005}. However this first interpretation had been revised after the observation of spin density wave (SDW) order in the HFLT phase by neutron diffraction \cite{Kenzelmann2008, Kenzelmann2010} and NMR experiments \cite{Young2007, Mitrovic2006, Koutroulakis2008, Koutroulakis2010}. The magnetic structure of  the $Q$ phase is indicated in Fig.~\ref{Kenzelmann}. Independent of the direction of the magnetic field in the basal plane, spin-density wave order with an incommensurate modulation \boldmath Q\unboldmath $=(q, q, 0.5)$ with $q = 0.44$ occurs with a small ordered magnetic moment of $\sim 0.15 \mu_B$ pointing along the $c$ direction, independent of the strength of the magnetic field. Indications of some exotic superconducting crossover phase (ESC), different from the usual Abrikosov vortex state but also from the $Q$-phase,  has been detected even below the second order phase transition to the $Q$-phase\cite{Koutroulakis2010}. These observations exclude a pure FFLO scenario for the HFLT phase. However, the fact that the HFLT phase is restricted to the superconducting state indicates the strong interplay between both orders. 
The origin of the field-induced SDW in CeCoIn$_5$ has led to the development of  several theoretical approaches. All of these are based on the closeness of the system to an antiferromagnetic quantum  critical point which leads to enhanced spin fluctuations, strong Pauli limit, and the possible appearance of an extra $\pi$ (triplet) superconducting component. This opens the possibility to mix singlet and triplet Cooper pairing channels via breaking the translation invariance when the HFLT phase  is established. 

In Ref.~\cite{Kenzelmann2008} it is discussed that the addition to the ambient d-wave superconducting component $\Delta_0$, the SDW order is coupled to a pair density wave (PDW). It has been suggested that 
the rapidly modulated SDW is coupled to $\pi$-pairing superconductivity of odd parity and arises from the presence of a long-wavelength modulated PDW associated with the Fulde-Ferrell-Larkin-Ovchinnikov (FFLO) state. 
Aperis $et$ $al.$~have pointed out that the HFLT state may represent a pattern of coexisting condensates: d-wave singlet superconducting (SC) state, a staggered $\pi$-triplet SC state, and a spin density wave (SDW) \cite{Aperis2010}. 
In the model of Yanase and Sigrist \cite{Yanase2008b, Yanase2009} the SDW order can arise in the FFLO state  as a consequence of the formation of Andreev bound states near the zeros of the FFLO order parameter. It is the large density of states in the bound states that triggers the formation of the incommensurate SDW order. 
Other models are based only on the strong Pauli paramagnetic superconductivity and the nodal gap structure of the $d_{x^2-y^2}$ symmetry \cite{Ikeda2010,  Suzuki2010, Kato2011}. In Ref.~\cite{Michal2011} an approach is given which connects the high field antiferromagnetism to the spin resonance.  

From an experimental point of view, a coexistence of the $Q$-phase with a FFLO one has not been clearly demonstrated. Future microscopic experiments, like neutron scattering or NMR may give further experimental grounds. E.g. in Refs.~\cite{Agterberg2009, Yanase2009} the appearance of additional satellite peaks in neutron scattering have been predicted, but up to now all experiments have failed to resolve such peaks. Another possibility will be to apply pressure, as e.g.~Ref.~\cite{Yanase2009} predicts a decoupling of an FFLO state and the field induced magnetism with increasing pressure. The response to impurities of SDW order and superconductivity may be different too. 

By turning the field from $H \parallel a$ to $H \parallel c$ the HFLT phase vanishes for angles above $\phi \approx 20^\circ$ \cite{Correa2007, Blackburn2010} and the vector of the SDW order does not change with angle ( at least up to 12$^\circ$). This angular dependence questions the connection of the anomalies seen close to the upper critical field $H_{c2}$ along the crystallographic $c$ axis. No evidence for a $Q$-phase exists for field along $c$ axis and the anomalies reported in NMR \cite{Kumagai2006} and $\mu$SR experiments \cite{Spehling2009} do not indicate a phase transition line from a vortex state to a FFLO or a magnetically ordered state. For $H \parallel c$ the superconducting transition gets first order too and the Pauli depairing is the dominant mechanism to suppress SC close to $H_{c2}$ \cite{Bianchi2008}. In Ref.~\cite{Suzuki2010} it has been stated that the appearance of the $Q$ phase will be restricted to the basal plane  where the lines nodes of the superconducting gap occur.

The phase diagram of CeCoIn$_5$ for $H$ along the $c$ axis has been studied mainly with the focus on  a putative field induced quantum critical point very close to $H_{c2}(0)$ and  the vortex lattice which is strongly connected to the Pauli limiting field. A putative field induced quantum critical point in CeCoIn$_5$ has been first discussed on the basis of transport measurements \cite{Paglione2003} and specific heat experiments \cite{Bianchi2003c} to coincide with the upper critical field $H_{c2}(0)$ indicating that the Fermi liquid regime collapses right at the upper critical field. In the specific heat a logarithmic increase to low temperatures has been observed for $H = 5$~T and for $H > 8$~T a Fermi liquid regime has been recovered. However, due to the strong hyper-fine contributions to the specific heat the very low temperature regime cannot be attended. Thermal conductivity experiments for $H\parallel c$ give no conclusive result on the break down of the Fermi liquid regime \cite{Paglione2006, Tanatar2007, Smith2008}. 

Recently resistivity experiments to much lower temperatures ($T > 8$~mK) show that the Fermi liquid regime does not collapse at $H_{c2}(0)$, but stays finite \cite{Howald2011}. An extrapolation from the paramagnetic regime would indicate a magnetic quantum critical field below $H_{c2}(0)$ at 0.92$H_{c2}(0)$, not far from the field where the superconducting transition at $T_c$ gets first order. A same putative quantum critical point inside the superconducting phase has been referred from Hall effect \cite{Singh2007} and recently from thermal expansion experiments \cite{Zaum2011}. 

Careful magnetization measurements allowing to extract both the magnetization and the field variation of the $\gamma$ coefficient below $H_{c2}$ shows no evidence of a HFLT phase for this direction \cite{Paulsen2011}. Furthermore the singularity of $\gamma$ occurs at $H_{c2}(0)$. Thus we want to point out that the appearance of magnetic criticality occurs through the interplay between magnetism and superconductivity and not at all through an isolated mechanism. This is in agreement with the theoretical result that superconductivity will enhance the proximity to a magnetic quantum critical point since it reverses the sign of the mode-mode coupling term in the spin-fluctuation frame \cite{Fujimoto2008}   The importance of the Pauli depairing is reflected in the strong increase of $\gamma (H)$ on approaching $H_{c2} (0)$. The novelty in CeCoIn$_5$ is that the field induced antiferromagnetism is pegged to $H_{c2}$ for $H \parallel a$ and never occurs for $H>H_{c2}$ or for $H \parallel c$. The link to antiferromagnetic quantum criticality induces static antiferromagnetic order for $H\parallel a$ due to the coherence of the Cooper pairs and in difference for $H\parallel c$ only an enhancement of the fluctuations towards a quantum critical point is realized.

\subsection{CeCoIn$_5$ under high pressure}

\begin{figure}[t]
\begin{center}
\includegraphics[width=0.8\hsize,clip]{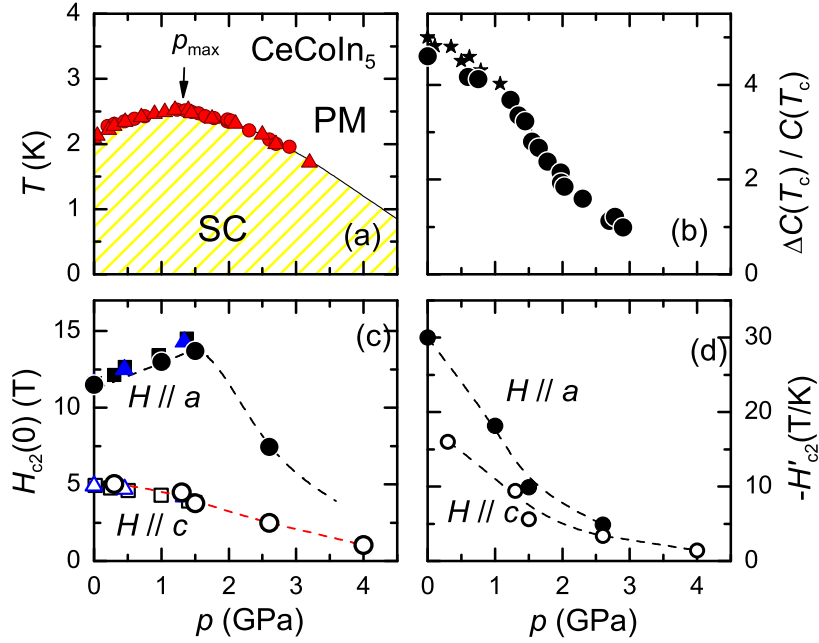}%
\caption{\label{CeCoIn5_PD} (a) High pressure phase diagram of CeCoIn$_5$ (from Ref.~\citealp{Knebel2004}). (b) Pressure dependence of the specific heat jump at the superconducting transition \cite{Knebel2004}. (c) Pressure dependence of the upper critical field $H_{c2} (0)$ for different crystallographic directions. (Circles from Ref.~\citealp{Knebel2010}, triangles from Ref.~\citealp{Miclea2006}, squares from  \citealp{Tayama2005}). (d) Pressure dependence of the initial slope $-H' = dH_{c2}/dT$ at $T_c$ for $H\parallel a$ and $H\parallel c$. 
} 
\end{center} 
\end{figure}

\begin{figure}[t]
\begin{center}
\includegraphics[width=0.5\hsize,clip]{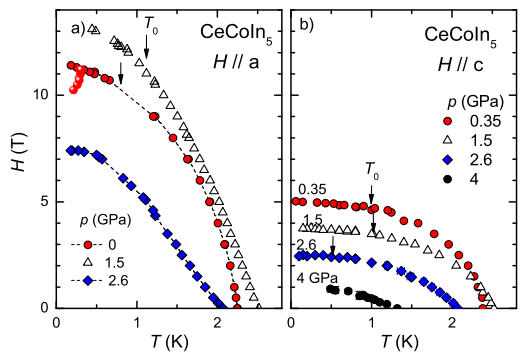}%
\caption{\label{CeCoIn5_Hc2_pressure} (a) Upper critical field of CeCoIn$_5$ for field along the $a$ axis and $c$ axis. The arrows indicate the field where the superconducting transition changes to first order. Under pressure, no signature of the HFLT phase could be observed, in difference to Ref.~\citealp{Miclea2006}. 
} 
\end{center} 
\end{figure}

The high pressure phase diagram of CeCoIn$_5$ is shown in Fig.~\ref{CeCoIn5_PD}. $T_c$ increases first under pressure up to $T_c \approx 2.6$~K at $p_{max} = 1.3$~GPa and the superconducting dome extends at least up to 5~GPa \cite{Sidorov2002, Knebel2004}. In magnetically mediated superconductivity it is generally believed that the maximum of $T_c$ appears just right at the antiferromagnetic critical point where the effective mass due to the fluctuation takes its maximum, as previously discussed for CeRhIn$_5$. However, the general picture for CeCoIn$_5$ is  that with increasing pressure antiferromagnetic fluctuations are suppressed and thus the systems is tuned away from the quantum critical point.  This has been clearly shown by NQR experiments \cite{Yashima2004} and also by dHvA experiments which show a decrease of all cyclotron masses \cite{Shishido2003}.

In Fig.~\ref{CeCoIn5_Hc2_pressure} we show the upper critical field $H_{c2}$ for different pressures determined by ac calorimetry \cite{Knebel2010}. The pressure dependence shows two remarkable points: (i) for $H \parallel a$ the upper critical field $H_{c2}(0)$ has its  maximum at $p_{max}$ where $T_c$ is maximal, but for $H \parallel c$, $H_{c2}(0)$ decreases with pressure even when $T_c$ increases, a fact which cannot be easily understood. As shown in Fig.~\ref{CeCoIn5_PD}(b) the size of the superconducting specific heat anomaly and the initial slope of the upper critical field (Fig.~\ref{CeCoIn5_PD}(d)) decreases monotonously with increasing pressure without any strong anomaly at $p_{max}$. From the initial slope it can be estimated that the effective mass decreases by almost an factor of two for both field directions, in agreement with the cyclotron masses determined in dHvA experiments \cite{Shishido2003}. 

We want to emphasize that in CeCoIn$5$ the variation of the upper critical field $H_{c2}(T)$ under pressures is not only determined by the strong coupling constant $\lambda$, by contrast to the case of CeRhIn$_5$ \cite{Knebel2008}. The main difficulty to be explained is the strange behavior for $H$ along the $c$ axis, where the Pauli limiting is strongest, as with increasing pressure up to 1.5~GPa $T_c$ is increasing, but $H_{c2}(0)$ is decreasing. 
In Ref.\cite{Kos2003} the possible coupling of the superconducting order parameter to fluctuating paramagnetic moments is discussed and it is shown that the presence of uncompensated moments give rise to a suppression of $T_c$ and an increase of the jump at the superconducting transition temperature. Taking such a paramagnetic pair breaking mechanism, which will decrease with pressure, into account, a qualitative explanation of the pressure dependence of the $H_{c2} (p)$ is given in Ref.~\cite{Howald2011b}. 

In our experiment we have not been able to follow the pressure dependence of the HFLT phase \cite{Knebel2010}. In  previous specific heat experiments \cite{Miclea2006} the HFLT phase has been found to expand under high pressure up to 1.5~GPa while magnetic fluctuations are suppressed with pressure. This has been interpreted as an indication that the HFLT phase may be coupled to an FFLO state and the conditions favorable for the formation of an FFLO state are still valid. E.g.~the first order nature of the superconducting transition can be followed up to 2.6~GPa along the $c$ axis and the Maki parameter, which is strongly pressure dependent,  decreasing from $\alpha = 4.4$ to 1.34 for $H \parallel a$ and from 7.4 to 1.8 for $H \parallel c$ axis. Of course, the microscopic evidence of the HFLT phase under high pressure is missing up to now and deserves new experiments in the future. E.g.~in Ref.~\cite{Yanase2009}
is has been proposed that under pressure, due to the suppression of the antiferromagnetic fluctuations a separation of the SDW order and a FFLO transition should be observed and thus two separate transition lines should be observed. However, this has not been observed in the specific heat experiment, but should be observable by e.g.~NMR experiments and small changes in the NMR spectra could be resolved.  Clearly, the stabilization of the $Q$-phase is a complex interplay between magnetism and superconductivity which involves the strong coupling regime and of the topology of the lines of zeros, i.e.~the dispersion of the Cooper pair excitations.

\section{Non-centrosymmetric Superconductors}

\begin{figure}[t]
\begin{center}
\includegraphics[width=0.4\hsize,clip]{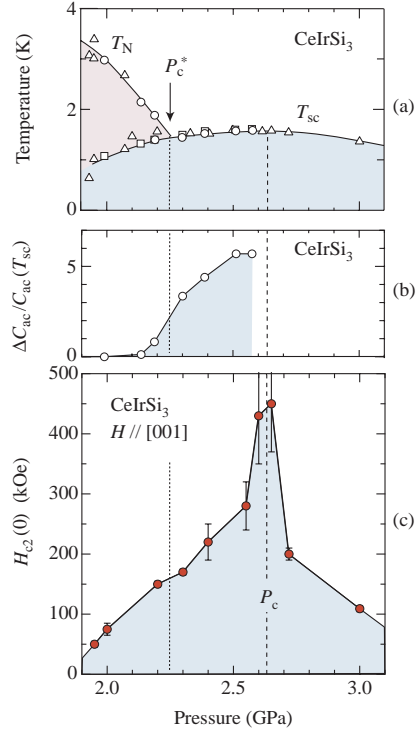}%
\caption{\label{CeIrSi3_PD} Pressure dependence of (a) the N\'eel temperature $T_N$ and the superconducting transition temperature $T_c$, (b) specific heat jump at the superconducting transition, and (c) the upper critical field $H_{c2}$ for $H \parallel$[001] in CeIrSi$_3$ (taken from Ref.~\cite{Settai2008}). 
} 
\end{center} 
\end{figure}

The crystal structure of all above discussed superconductors are characterized by a center of inversion symmetry which allows a classification in even (spin singlet) or odd parity (spin triplet) superconducting order parameters \cite{Anderson1984}. Thus, the observation of superconductivity at $T_c\approx 0.7$~K in non-centrosymmetric CePt$_3$Si below the antiferromagnetic transition at $T_N = 2.2$~K has been a surprise and opened another very rich field of intense research \cite{Bauer2004}. In strongly correlated electron systems there exist now different interesting cases: CePt$_3$Si \cite{Bauer2004, Bauer2007}, and the members of the so-called Ce-113 family CeRhSi$_3$ \cite{Kimura2005}, CeIrSi$_3$ \cite{Sugitani2006, Settai2007}, CeCoGe$_3$ \cite{Settai2007,Knebel2009} and CeIrGe$_3$ \cite{Honda2010}. For all examples superconductivity appears in the pressure window where the antiferromagnetism will be suppressed (see e.~g.~Fig.\ref{CeIrSi3_PD} (a)). Furthermore, as in CeRhIn$_5$ at $H =0$, above $p_c^\star$, where $T_N(p) = T_c(p)$, the superconducting phase transition as measured by resistivity or specific heat is very sharp and the phase diagrams are very similar. 

In addition to the strong correlations due to the antiferromagnetic instability, another key input is the strong anisotropic spin-orbit interaction caused by the lack of inversion symmetry. This Rashba-like type of interaction causes a lifting of degenerate bands in the density of states, separating spin up and spin down states. If the energy of spin-orbit coupling is much larger than the superconducting gap, interband Cooper pairing between electrons of the spin-orbit split bands is prevented which gives new consequences, e.g.~mixing of even and odd pairing superconducting states are now allowed \cite{Frigeri2004a, Sigrist2007, Fujimoto2007a}. This mixing can give lines in quasi-particle excitations. 

Here we will not review all aspects of non-centrosymmetric systems, recently detailed reviews are given in Refs.~\cite{Bauer2007,Onuki2011}. Let us focus on the high pressure, high magnetic field phase diagram of CeIrSi$_3$ \cite{Settai2008} which is shown in Fig.~\ref{CeIrSi3_PD}. Antiferromagnetic order is observed up to $p_c^\star$, without superconductivity the magnetic order would vanish at the critical pressure $p_c > p_c^\star$. For $p < p_c^\star$ a coexistence regime of magnetism and superconductivity exist, but e.g.~the specific heat anomaly at the superconducting transition below $p_c^\star$ is very small in zero field. When entering in the pure superconducting regime above $p_c^\star$ the jump of $\Delta C_{ac} / C_{ac}$ at $T_c$ reaches a broad maximum when $T_c(p)$ goes through a broad maximum \cite{Tateiwa2007}. 

The surprise here is the huge value for $H_{c2}(0)$ close to the pressure $p_c$ where $T_c$ has its maximum indicating that the system reaches some criticality under magnetic field right at $H_{c2}(0)$ for $p = p_c$. 
This phenomena seems now well explained in a formulation where the Pauli and orbital depairing effect are treated on equal footing with an interplay between the Rashba spin-orbit interaction and the spin fluctuations enhanced near the quantum critical point \cite{Tada2008,Tada2010}. This can also clarify the large anisotropy of the upper critical field between $H \parallel c$  where no Pauli limitation occurs by contrast to $H \perp c$. We want to notice that no field re-entrance of antiferromagnetism has been observed in the pressure range between $p_c^\star < p < p_c$ \cite{Settai2011}.

\section{Conclusion}

In this article we concentrated on the $(T,p,H)$ phase diagram of Ce-based heavy fermion superconductors. The interplay between antiferromagnetic order and superconductivity associated with the proximity to a magnetic quantum phase transition gives strong evidence that the unconventional superconductivity is generated by spin and/or valence fluctuations. Both instabilities often cannot be distinguished and are mixed. The interest of heavy fermion systems are the chance to be able to grow excellent large single crystals and the  opportunity to perform a large variety of experiments in a parameter space which is comfortable for experimentalists: a large temperature window due to high ordering temperatures $T_N \sim T_c \sim 2$~K, moderate magnetic fields $H_{c2} \sim 10$~T, and comfortable pressure range $p < 3.5$~GPa for reliable microscopic and macroscopic probes. This allowed rapid breakthroughs notably in the Ce-115 family. 

The exotic properties are the result of strong electronic correlations, the proximity to a magnetic and/or a valence instability, and the specificity of unconventional superconducting ground states characterized by their line or point nodes. For experimentalists and for theoreticians, it is of course challenging to measure, predict or analyze what will be the response in the pressure--field parameter space. Many observations were unexpected as very often only experimental facts lead to realize the originality of the situation. The heavy fermion materials have allowed to observe very often first unique phenomena with great accuracy. The final message is with positive mind, lucky interesting, and unexpected results can emerge. 


\section*{Acknowledgements} 
We thank J.-P.~Brison, D.~Braithwaite, A.~Demuer, H.~Harima, L.~Howald, K.~Izawa, D.~Jaccard, W.~Knafo, G.~Lapertot, M.-A.~Measson, K.~Miyake, S.~Raymond, E.~Ressouche, B.~Salce, R.~Settai, and I.~Sheikin for fruitful discussions. 
We acknowledge financial support to this work from the French National Research Agency (ANR with the contracts CORMAT, DELICE and ECCE), the European Research Council (ERC junior NewHeavyFermion). JF is supported as "director de recherche emeritus" by CNRS.










\end{document}